\documentclass[12pt,letterpaper,usenames,dvipsnames]{article}
\usepackage{jheppub}

\allowdisplaybreaks[2] 
\usepackage{bm} 
\usepackage{dsfont} 
\usepackage{bbm} 
\usepackage{mathrsfs} 
\usepackage{stmaryrd} 
\usepackage{physics} 
\usepackage{outlines} 
\usepackage[normalem]{ulem} 

\usepackage{tikz}
\usetikzlibrary{cd,arrows,calc,shapes,shapes.arrows,
decorations.markings,decorations.pathmorphing}

\usepackage{multirow} 
\usepackage{colortbl} 
\usepackage{arydshln} 

\usepackage{xcolor}

\definecolor{purple}{HTML}{9B00FF}

\definecolor{tred}{HTML}{E00122}
\def\tred#1{{\color[HTML]{E00122}{#1}}}
\definecolor{gfruit}{HTML}{ff5b59}

\definecolor{rem}{HTML}{0097AD}

\definecolor{seagreen}{HTML}{009382}

\definecolor{water}{HTML}{007AFF}

\definecolor{comp}{HTML}{5B68FF}

\definecolor{dblue}{HTML}{014FFF}

\definecolor{lblue}{HTML}{00C7E9}
\definecolor{teal}{HTML}{0097AD}

\usepackage{todonotes}

\def\nn{\nonumber}
\newcommand{\bpm}{\begin{pmatrix}}
\newcommand{\epm}{\end{pmatrix}}
\newcommand{\bsm}{\begin{smallmatrix}}
\newcommand{\esm}{\end{smallmatrix}}

\newcommand{\bspm}{\left(\begin{smallmatrix}}
\newcommand{\espm}{\end{smallmatrix}\right)}
\newcommand{\nord}[1]{{:}\mkern1mu#1\mkern1.6mu{:}} 
\newcommand{\deq}{\doteq}
\def\<{{\langle}}
\def\>{{\rangle}}

\def\bar{\overline}
\def\til{\widetilde}
\def\wh{\widehat}
\def\dag{\dagger}
\def\^{\wedge}
\def\del{{\partial}}

\def\Re{{\rm Re}}
\def\sgn{{\rm sgn}}

\def\sl{\mathfrak{sl}}

\def\SO{{\rm SO}}

\def\SU{{\rm SU}}
\def\su{\mathfrak{su}}

\def\U{{\rm U}}

\def\fB{{\mathfrak B}}
\def\C{\mathbb{C}} 
\def\bD{{\mathbf D}}

\def\cfE{{\mathscr E}}
\def\cfF{{\mathscr F}}
\def\hg{{\wh g}}

\def\cfI{{\mathscr I}}

\def\tj{{\til\jmath}}
\def\ellb{{\bar\ell}}
\def\hL{{\wh L}}
\def\bL{{\bf L}}
\def\bLm{{\bf L^-}}
\def\bLp{{\bf L^+}}
\def\cM{{\mathcal M}}
\def\cMt{{\til\cM}}
\def\fm{{\mathfrak m}}
\def\N{\mathbb{N}} 
\def\cN{{\mathcal N}}
\def\cO{{\mathcal O}}
\def\P{\mathbb{P}} 
 
\def\QQ{{\mathbb Q}}
\def\Qt{{\til Q}}
\def\R{\mathbb{R}} 
\def\St{{\til S}}
\def\bS{{\bf S}}
\def\hbS{{\wh\bS}}
\def\fS{{\mathfrak S}}
\def\hS{{\wh S}}
\def\TT{{\mathbb T}}
\def\trU{\tred{U}}
\def\trV{\tred{V}}

\def\bW{{\bf W}}
\def\hW{{\wh W}}
\def\hbW{{\wh\bW}}
\def\bX{{\mathbf X}}
\def\bY{{\mathbf Y}}
\def\cZ{{\mathcal Z}}
\def\Z{\mathbb{Z}}
\def\fZ{{\mathfrak Z}}
\def\zb{{\bar z}}
\def\a{{\alpha}}
\def\ad{{\dot\a}}
\def\b{{\beta}}
\def\bb{{\bar\b}}
\def\bd{{\dot\b}}

\def\d{{\delta}}
\def\D{{\Delta}}
\def\e{{\epsilon}}

\def\z{{\zeta}}
\def\th{{\theta}}
\def\k{{\kappa}}

\def\m{{\mu}}
\def\bm{{\boldsymbol\mu}}
\def\n{{\nu}}
\def\x{{\xi}}
\def\r{{\rho}}
\def\s{{\sigma}}
\def\S{{\Sigma}}

\def\f{{\phi}}

\def\ps{{\psi}}
\def\pst{{\til\ps}}

\title{Vertex algebra of extended operators in 4d N=2 superconformal field theories}
\author[a]{Philip C. Argyres,}
\author[b]{Matteo Lotito,}
\author[a]{Mitch Weaver}
\affiliation[a]{Physics Department, University of Cincinnati, PO Box 210011, Cincinnati OH 45221 USA}
\affiliation[b]{Department of Physics and Astronomy \& Center for Theoretical Physics,
Seoul National University, Seoul 08826, Republic of Korea}
\emailAdd{philip.argyres@gmail.com}
\emailAdd{matteolotito@gmail.com}
\emailAdd{weaverm5@mail.uc.edu}

\abstract{
We construct a class of extended operators in the cohomology of a pair of twisted Schur supercharges of 4d $\cN{=}2$ SCFTs.  
The extended operators are constructed from the local operators in this cohomology --- the Schur operators --- by a version of topological descent. 
They are line, surface, and domain wall world volume integrals of certain super descendants of Schur operators.  
Their world volumes extend in directions transverse to a spatial plane in Minkowski space-time. 
As operators in the cohomology of these twisted Schur supercharges, their correlators are (locally) meromorphic functions only of the positions where they intersect this plane. 
This implies the extended operators enlarge the vertex operator algebra of the Schur operators.
We illustrate this enlarged vertex algebra by computing some extended-operator product expansions within a subalgebra of it for the free hypermultiplet SCFT. 
}

\begin{document}
\setcounter{tocdepth}{2}
\maketitle


\section{Extended operators in twisted Schur cohomology}
\label{sec:1}

Extended operators, such as Wilson-'t Hooft line operators, play a central role in our understanding of gauge field theories.
Progress on analytic control of correlation functions of extended operators in strongly-coupled gauge theories has been made over the past 30 years by examining extended operators in supersymmetric gauge theories and the closely related constructions of topological extended operators in topologically twisted supersymmetric gauge theories.
But the spectrum and correlators of supersymmetric extended operators in strongly coupled non-lagrangian supersymmetric field theories --- e.g., in isolated superconformal field theories (SCFTs) --- is much less well understood.


In the last decade much insight into the structure of the local operator algebra of 4d $\cN{=}2$ SCFTs has come from the realization \cite{Beem_2015} that the operator algebra of certain BPS local operators --- the twist-translated Schur operators --- have the structure of a vertex operator algebra (VOA), which is more or less the chiral half of a 2d CFT. 
Twist-translated Schur operators are the local operators in the cohomology of a pair of nilpotent supercharges in the superconformal algebra, $\TT_\pm$, called twisted Schur supercharges, and their operator algebra is the VOA.

It is natural to ask, in this context, whether there are extended operators in twisted Schur cohomology, and if so, what is the structure of their operator algebra. 
We will show that there are, indeed, line, surface, and domain wall operators in twisted Schur cohomology.
They are constructed from the local Schur operators in a way that is analogous to how extended topological operators are constructed from local operators by topological descent \cite{Witten_1988}. 
However, unlike the topological extended operators constructed in \cite{Witten_1988}, we will see that the extended operators constructed from descent in twisted Schur cohomology are not topological.%
\footnote{Our surface operators are different from previously constructed examples of surface operators in twisted-Schur cohomology \cite{Cordova:2017,Beem_2018,Bianchi_2019}.}
The world volumes of the line operators and two of the surface operators are supported only on light-like submanifolds of a light cone, and so only exist in Lorentzian signature space-time. 
Furthermore, we will show that their correlators, with each other and with twist-translated Schur operators, depend only on a single complex coordinate, $z$, entering into the parameterization of the space-time position of each extended operator insertion.
The $(z,\bar z)$ coordinates parameterize a spatial 2-plane in space-time which we will call the VOA plane.
Thus, the combined operator algebra generated by the extended operators and the twist-translated Schur operators expands the original VOA to a vertex algebra.

This note provides a concise introduction to and summary of these results. 
A more extensive derivation, including many technical details, is provided in \cite{Argyres_2023xx}.

\paragraph{Summary of results.}

VOAs corresponding to 4d $\cN{=}2$ SCFTs arise when one studies the protected sector of Schur operators annihilated by the twisted Schur supercharges introduced in \cite{Beem_2015}.
We will consider the pair of nilpotent twisted Schur supercharges
\begin{align}\label{Tpmdef}
\TT_+ &\doteq Q^1_{~2} + \St^{2\dot2} , &
\TT_- &\doteq S_1^{~2}  - \Qt_{2\dot2} .
\end{align}
We follow the notation and conventions of \cite{Lemos_2020} unless otherwise specified.%
\footnote{We use $1$ and $2$ for spinor indices in place of $+$ and $-$ used in \cite{Beem_2015, Lemos_2020}.  We reserve $\pm$ for lightcone indices.  $\TT_\pm$ are the same as $\QQ_1$ and $\QQ_2$ in \cite{Beem_2015} and are those of \cite{Lemos_2020} with phase $\z=1$.}
Local operators in $\TT_+$-cohomology are automatically also in $\TT_-$-cohomology in a unitary SCFT.

The set of extended operators that we analyze in this work are of a special type, chosen in order to guarantee that they are non-trivial objects in the cohomology of the twisted Schur supercharges.
Starting from a local Schur operator, we perform a version of topological descent \cite{Witten_1988} by $\TT_\pm$-exact generators of the conformal algebra.
This gives rise to a web of descent operators obtained from a single local Schur operator, $\cO$,
\begin{align}
\cO \to \bD^{X_1}[\cO] \to \bD^{X_2}\bD^{X_1}[\cO] \to \cdots .
\end{align}
Here $\bD^X$ denotes the descent operation corresponding to the exact generator $X$.

While there is a continuum of exact generators with which to perform descent, by combining $\TT_\pm$-cohomology equivalences between various descent operations with the requirement that descent operators be finite, the set of possible descent operators is reduced to a finite set or web.
Within this web, each successive descent increases the world volume dimension of the descent operator by one.
Each $\bD^X$ acts on the preceding operator by a (fermionic) supercharge, so flips the statistics of the descent operator;
each $\bD^X$ changes the $\U(1)_r$ charge by $\pm1$;
$\bD^X$ preserves the chiral weight of $\cO$, $h_\cO$, under the holomorphic conformal $\sl_2$ symmetry of the VOA plane; 
and $\bD^X$ commutes with twist translations in the sense that  $\bD^X[\del_z\cO] =_\TT \del_z\bD^X[\cO]$, so $\del_z$ still acts as the translation generator of the extended vertex algebra.
Here ``$=_\TT$'' means equal in $\TT_\pm$ cohomology.

\begin{figure}[ht]
\centering
\begin{tikzpicture}
[baseline={(current bounding box.center)},
 bac/.style={line width=7pt, white, shorten >=10pt},
 arr/.style={->,shorten >=3pt,>=stealth',thick}]
\node at (-8,0) {point};
\node at (-8,-2) {lines};
\node at (-8,-4) {surfaces};
\node at (-8,-6) {walls};
\node (Sc) at (0,0) {$\cO$};
\node (Lm) at (-1.7,-2) {$\bLm[\cO]$};
\node (Lp) at (1.7,-2) {$\bLp[\cO]$};
\node (S) at (0,-4) {$\bS[\cO]$};
\node[color= tred] (Stb) at (3.4,-4) {$\hbS_R[\cO]$};
\node[color= tred] (Sta) at (-3.4,-4) {$\hbS_L[\cO]$};
\node[color= tred] (Wmb) at (1.7,-6) {$\hbW_R^-[\cO]$};
\node[color= tred] (Wpa) at (-1.7,-6) {$\hbW_L^+[\cO]$};
\node[color= tred] (Wmpm) at (4.2,-6) {$\hbW_{R\pm}^+[\cO]$};
\node[color= tred] (Wppm) at (-4.2,-6) {$\hbW_{L\pm}^-[\cO]$};
\node[color= tred] (Wpb) at (6,-6) {$\hbW_R^+[\cO]$};
\node[color= tred] (Wma) at (-6,-6) {$\hbW_L^-[\cO]$};
\draw[arr] (Sc) to node [xshift=-7pt,yshift=7pt,font=\small] {$P_-$} (Lm);
\draw[arr] (Sc) to node [xshift=7pt,yshift=7pt,font=\small] {$P_+$} (Lp);
\draw[arr] (Lm) to node [xshift=7pt,yshift=7pt,font=\small] {$P_+$} (S);
\draw[arr, tred] (Lm) to node [xshift=-10pt,yshift=7pt,font=\small] {$M_{+3}$} (Sta);
\draw[arr] (Lp) to node [xshift=-7pt,yshift=7pt,font=\small] {$P_-$} (S);
\draw[arr, tred] (Lp) to node [xshift=9pt,yshift=7pt,font=\small] {$M_{-3}$}(Stb);
\draw[arr] (Sta) to node [xshift=7pt,yshift=7pt,font=\small] {$P_+$}(Wpa);
\draw[arr] (Sta) to node [xshift=14pt,yshift=-5pt,font=\tiny] {$P_-{\pm}K_-$}(Wppm);
\draw[arr] (Sta) to node [xshift=-7pt,yshift=7pt,font=\small] {$P_-$}(Wma);
\draw[arr, tred] (S) to node [xshift=-10pt,yshift=7pt,font=\small] {$M_{+3}$} (Wpa);
\draw[arr, tred] (S) to node [xshift=9pt,yshift=7pt,font=\small] {$M_{-3}$}(Wmb);
\draw[arr] (Stb) to node [xshift=-7pt,yshift=7pt,font=\small] {$P_-$} (Wmb);
\draw[arr] (Stb) to node [xshift=-14pt,yshift=-5pt,font=\tiny] {$P_+{\pm}K_+$} (Wmpm);
\draw[arr] (Stb) to node [xshift=7pt,yshift=7pt,font=\small] {$P_+$} (Wpb);
\end{tikzpicture}
\caption{ 
The descent web for a general Schur operator $\cO$. Descent operations $\bD^X$ are shown as arrows labeled by the generator $X$. 
The resulting descent operator cohomology classes are labeled by $\bL^{\pm}$, $\bS$, $\hbS_{\dots}$, $\hbW^\pm_{\dots}$ which have line, surface, and domain wall world volumes. 
For example, $\bS[\cO] \doteq \bD^{P_-}\bD^{P_+}[\cO] = -\bD^{P_+}\bD^{P_-}[\cO]$.  
The $\pm$ superscripts denote $\U(1)_r$ charges of the corresponding descent procedure, while $L$, $R$ subscripts denote the chirality of the action of $M_{\pm 3}$ on $\cO(0)$.
The hatted descent operators are shown in red because, as discussed in sections \ref{descent_web} and \ref{hat_conv}, they have exotic properties, and there may be equivalences between some of their cohomology classes not shown in the figure.
}
\label{web}
\end{figure}

The resulting descent web is shown in figure \ref{web}.
The arrows are the descent operations labeled by their corresponding generators.
$P_\pm$ are certain light-like translations and $M_{\pm3}$ are light-like combinations of boosts and rotations, defined below.
These generators also determine the space-time world volumes of the descent operators; they are illustrated in figure \ref{fig1} below.
We have introduced shorthand names, $\bL^\pm$, $\bS$, $\hbS_{\dots}$, $\hbW^\pm_{\dots}$, for the iterated descent operations.
The $\pm$ superscripts are correlated with the $\U(1)_r$ charges.
The statistics of the descent operators alternate between the rows of figure \ref{web}, while their chiral weights are all the same.
This is reminiscent of the structure of the chiral half of a 2d topological SCFT.
But, as will become evident later, it is not at all clear whether the enlarged vertex algebra has the structure of a superconformal vertex algebra.

The $\hbS_{\dots}$ and $\hbW^\pm_{\dots}$ extended surface and wall operators, which we will often refer to as ``hatted descent operators" and are shown in red in figure \ref{web}, can only be reached via descent with an $M_{\pm3}$ generator.
These descent operators have the distinct feature that their correlators have a sign discontinuity when other vertex operator insertions cross a quasi-topological line emanating from them in the VOA plane.
Unlike local Schur operators, this behavior means they are not, strictly speaking, vertex operators in the vertex algebra, but suggests they should be viewed as ``boundary" operators.%
\footnote{$\hbW^{\pm}_{\ldots}$ operators can literally be interpreted as boundaries in the VOA plane because their world volumes intersect the VOA plane along a line; see figure \ref{fig1}. 
This line of intersection always sits orthogonal to the quasi-topological ``cut" all correlators of hatted operators possess, so is also quasi-topological. 
We expect this physical intersection within the VOA plane to introduce additional ``intersection" discontinuities in correlators of $\hbW^{\pm}_{\ldots}$ descent operators.}
We can describe this situation by saying that the insertion of a hatted operator introduces a ``cut'', or ``cuts", in the VOA plane.
In this paper, we describe many aspects of these interesting features of the hatted descent operators, but only present explicit OPE computations for the subalgebra formed from the $\cO$-$\bL^\pm$-$\bS$ operators, shown in black in figure \ref{web}.

All descent operators require a regularization prescription to ensure that their correlators are absolutely convergent and satisfy the relevant Ward identities defining a consistent $\TT_{\pm}$-cohomology theory.
We propose such a prescription which involves modifying the naive definition of a descent operator via topological descent to include a ``weight function".
These weight functions comprise a class of functions that satisfy specific properties which ensure descent operator correlators are absolutely convergent and obey all requisite $\TT_{\pm}$ Ward identities.
We also present preliminary evidence suggesting this prescription will produce universal, i.e., weight function independent results, but we do not have a general proof demonstrating that this is always true.
Additionally, the regularization prescription may obscure possible cohomological equivalences among descent operators at the level of the vertex algebra. 

In the next section, we define the twisted Schur descent operation and derive the descent web, figure \ref{web}.
In section \ref{sec:3}, we first discuss the convergence and regularization of descent operators. 
This is followed by an examination of the world volume intersections and intricate $\TT_{\pm}$ Ward identity discontinuities that can exist in multi-descent correlators.
Then, in section \ref{sec:4}, we compute some operator product expansions in the $\cO$-$\bL^\pm$-$\bS$ vertex subalgebra for the free hypermultiplet SCFT.
We finish in section \ref{sec:5} with remarks on future directions and open problems.


\section{Twisted Schur descent}\label{sec:2}

\subsection{Topological descent}

Recall the topological descent procedure \cite{Witten_1988} for constructing extended operators from point operators in topological field theories (TFTs) formed by twisting supersymmetric theories by a nilpotent supercharge $\QQ$.
In a TFT the translation generators $P_\m$ are $\QQ$-exact, so 
there exist fermionic generators $Q_\m$ such that%
\footnote{We use $\circ$ to denote the adjoint action for superalgebra and group elements.  Thus $\circ$ is the (anti)commutator between algebra generators, and also denotes conjugation of a generator by a group element.  We use the convention that when there are no parentheses $\circ$ acts on everything to its right.}
\begin{align}\label{TD1}
\QQ\circ P_\m &=0, & &\text{and}& 
P_\m &= \QQ\circ Q_\m .
\end{align} 
If $\cO(0)$ is a $\QQ$-closed local operator, then so is $\cO(x) \doteq  e^{x^\m P_\m} \circ \cO(0)$.
Define the extended operator
\begin{align}\label{TD3}
{\bf \S}[\cO] \doteq  \int_\S 
(Q\cdot dx) \circ \cdots \circ (Q\cdot dx) \circ \cO(x),
\end{align}
where there are $p$ factors of $Q\cdot dx \doteq Q_\m dx^\m$ in the integrand, and $\S$ is some closed $p$-manifold.  ${\bf\S}[\cO]$ is $\QQ$-closed since the $\QQ$-variation of the integrand is a total derivative by \eqref{TD1}.
Furthermore, the integrand is a closed $p$-form in $\QQ$-cohomology, so the $\QQ$-cohomology class of ${\bf\S}[\cO]$ only depends on the homology class of $\S$.  These topological properties of descent operators follow from 
\begin{align}\label{TD4}
Q_\m \circ Q_\n = 0, \qquad \forall\  \m,\n,
\end{align}
which is an additional property of $\QQ$ in a TFT.

\subsection{Generalized descent}

Recast topological descent as an iterative process, where, starting with a perhaps extended operator $\cfE$ in $\QQ$-cohomology, we define its descent with respect to the exact generator $n\cdot P \doteq n^\m P_\m$, where $n^\m$ is a unit vector, as the new operator extended in the $n^\m$ direction by an operation
\begin{align}\label{GD1}
 \bD^{n\cdot P} [\cfE] \doteq \int d\a\, (n\cdot Q) \circ e^{\a \, n\cdot P} \cfE .
\end{align}
That $\bD^{n\cdot P} [\cfE]$ is in $\QQ$-cohomology and that $[\bD^{n\cdot P},\bD^{m\cdot P}]=0$ follow from \eqref{TD4}.

Let us generalize this construction of extended operators in cohomology to any nilpotent supercharge $\TT$ in a superconformal algebra.  For any $\TT$-exact real bosonic generator $X$, there is a supercharge $\x$ such that 
\begin{align}\label{GD2}
X=\TT\circ\x .  
\end{align}
Define the $X$-descent of any (perhaps extended) $\TT$-closed operator $\cfE$ as
\begin{align}\label{GD3}
\bD^X[\cfE] \doteq  \int\! d\a\, \x\circ e^{\a X} \circ \cfE ,
\end{align}
where the integral is over the real line or a circle if the 1-parameter subgroup generated by $X$ is non-compact or compact, respectively.   In all the cases of interest to us it will be non-compact, so from now on $\int d\a = \int_{-\infty}^\infty d\a$.   Since $X$ is real, $e^{\a X}\circ \cfE$ is a conformally transformed operator supported at the conformal transform of its original space-time support.  
Then $\bD^X[\cfE]$ is $\TT$-closed up to boundary terms,
\begin{align}\label{GD4}
\TT\circ \bD^X[\cfE] &= \int\! d\a \, \TT\circ \x\circ e^{\a X} \circ \cfE
= \int d\a \left\{ (\TT\circ \x)\circ e^{\a X} \circ \cfE -
\x\circ e^{\a X} \circ (e^{-\a X} \circ \TT) \circ \cfE \right\} \nn\\
&= \int d\a \left\{ X\circ e^{\a X} \circ \cfE -
\x\circ e^{\a X} \circ \TT \circ \cfE \right\} 
= \int d\a \frac d{d\a} \left\{ e^{\a X} \circ \cfE \right\} \nn\\
&= \Big [ e^{\a X} \circ \cfE \Big]_{\a=-\infty}^{\a=+\infty} = 0 \qquad \text{if} \qquad \lim_{\a\to\pm\infty} e^{\a X} \circ \cfE = 0.
\end{align}
We test whether ``$\lim_{\a\to\pm\infty} e^{\a X} \circ \cfE = 0$'' by asking if the left side gives zero when inserted into the integrand of $\TT$ Ward identities.
When the $\cfE$ are integrals of operators of sufficiently positive conformal dimension, and to the extent that $\lim_{\a\to\pm\infty} e^{\a X} \circ \cfE$ moves the operator support to space-time infinity, then in euclidean space, these boundary limits converge to zero in correlators.
But this convergence does not hold at light-like infinity in Minkowski space-time for multi-descent correlators.  As a result, a regularization for descent operators is required for $\TT$-cohomology Ward identities to be satisfied. This regularization is presented in section \ref{sec:3}.

From \eqref{GD4}, it similarly follows that if $\cfE = \TT\circ \cfF$ is $\TT$-exact, then its descent operators are also $\TT$-exact,
\begin{align}\label{GD5}
    \bD^X[\TT\circ \cfF] 
    &= -\TT\circ \bD^X [\cfF] ,
\end{align}
again only up to boundary terms which must be checked to vanish.
Using the $\TT$-closedness of $X$ and $\cfE$, it is then easy to see that the $\TT$-cohomology class of $\bD^X[\cfE]$ does not depend on the specific representative $\x$ chosen to solve \eqref{GD2}.

If $\cO$ is a point operator, say at the origin, then $\bD^X[\cO]$ is a line operator if $e^{\a X}$ moves the origin; otherwise it will be a new point operator.   
Similarly, when $\cfE$ is an extended operator localized on some $p$-dimensional variety $\S$, $\bD^X [\cfE]$ is either a $(p+1)$-dimensional extended operator if $e^{\a X}$ does not preserve $\S$ as a set, or is another $p$-dimensional extended operator if it does.  
In our case, descents which preserve $\S$ all either vanish or diverge, so only descents which increase the dimensionality of the operator support are interesting.
Note that the analog of the relations \eqref{TD4} which ensure the topological nature of the dependence of descent operators on their $\S$ world volumes in TFTs need no longer hold in the case of generalized descent.

The number and nature of the resulting $\TT$-cohomology classes of descent operators depends on the details of the algebra of the real bosonic $\TT$-exact generators.  

\subsection{Twisted Schur cohomology of the superconformal algebra}

We now specialize to descent with respect to the pair $\TT_\pm$ of twisted Schur supercharges defined in \eqref{Tpmdef}.
Unitarity ensures that point operators in $\TT_+$-cohomology are automatically also in $\TT_-$-cohomology; these are the Schur operators of the SCFT.
It is a non-obvious fact that this continues to hold for the extended operators constructed from the Schur operators by $\TT_\pm$ descent.
Thus, if $\cO$ is a Schur, and $\bD^+[\cO]$ is a descent operator by some $\TT_+$-exact generator, then by \eqref{GD4} $\bD^+[\cO]$ will be $\TT_+$-closed (up to boundary terms which must, of course, be shown to vanish).  
The non-obvious fact is that $\bD^+[\cO]$ is also $\TT_-$-closed, even though it is not found by descent with respect to any $\TT_-$-exact generator.

To show this and other special properties of twisted Schur descent, we compute the subalgebras of $\TT_\pm$-closed or exact bosonic generators of the $\cN{=}2$ superconformal algebra.
First, it is convenient to fix a space-time coordinate system.  
Let $x^\m$,~$\m~=~1,2,3,4$, be the usual flat Minkowski coordinates with $x^1$ the time, and define lightcone coordinates in the $x^1$-$x^2$ and complex coordinates in the $x^3$-$x^4$ planes by
\begin{align}
    x^\pm &\doteq \pm x^1 +x^2, &
    z &\doteq x^3 - i x^4 .
\end{align}
Points at $x^\pm=0$ with coordinates $(z,\zb)$ comprise the ``VOA plane".
The tensor generators of the conformal algebra, $P_\m$, $M_{\m\n}$, and $K_\m$, are related to their spinor components by%
\footnote{This differs slightly from the convention in \cite{Lemos_2020} because we are working in Minkowski space-time.}
\begin{align}
\bpm P_{1\dot1} & P_{1\dot2} \\  P_{2\dot1} & P_{2\dot2} \epm
&= \bpm 2P_z & -2iP_- \\ 2i P_+ & -2P_\zb \epm ,&
\bpm K^{\dot11} & K^{\dot12} \\  K^{\dot21} & K^{\dot22} \epm 
&=  2 \bpm K_\zb & -i K_- \\ i K_+ & -K_z \epm , \\
\bpm \cM_1^{\ 1} & \cM_1^{\ 2} \\  \cM_2^{\ 1} & \cM_2^{\ 2} \epm
&= \bpm  
-\cM_2^{\ 2} & 2iM_{-z}\\ 2iM_{+\zb} & M_{+-} {-} M_{z\zb}\epm ,&
\bpm \cMt^{\dot1}_{\ \dot1} & \cMt^{\dot1}_{\ \dot2} \\  
\cMt^{\dot2}_{\ \dot1} & \cMt^{\dot2}_{\ \dot2} \epm
&= \bpm 
-\cMt^{\dot2}_{\ \dot2} & 2iM_{-\zb}\\ 2iM_{+z}& M_{+-} {+} M_{z\zb}
\epm .\nn
\end{align}

Denote the subalgebras of the $\cN{=}2$ superconformal algebra $\fS$ which are closed and exact under twisted Schur supercharges $\TT_\pm$ by $\fZ_\pm \doteq \{ X\in\fS \ |\  \TT_\pm\circ X=0 \}$ and $\fB_\pm \doteq \{ X\in\fS \ |\  X=\TT_\pm \circ Y \text{ for some }Y\in\fS \}$, respectively.
Then the even (bosonic) subalgebras are%
\footnote{As indicated in \eqref{Tcohom3}, the $\fm_{\pm}$ subalgebras do not commute with the $\sl_2 \oplus \wh \sl_2 \oplus \cZ$ subalgebra, which we denote using the ``semi-direct sum" symbol $\niplus$.}
\begin{align}\label{Tcohom1}
    \fZ_\pm &= \sl_2 \oplus \wh\sl_2 \oplus \cZ \niplus \fm_\pm, &
    \fB_\pm &= \wh\sl_2 \oplus \cZ \niplus \fm_\pm ,
\end{align}
where
\begin{align}\label{Tcohom2}
\sl_2 &\doteq  \big\langle \ L_{-1} \doteq  P_z ,\qquad\quad
L_0 \doteq  \tfrac12 D + M_{z\zb} ,\qquad\quad
L_1 \doteq  K_\zb \ \big\rangle ,\\
\wh\sl_2 &\doteq  \big\langle \ \hL_{-1} \doteq  P_\zb - R^- ,\ 
\hL_0 \doteq  \tfrac12 D - M_{z\zb} - \tfrac12 R ,\ 
\hL_1 \doteq  K_z + R^+ \ \big\rangle ,\nn\\
\cZ &\doteq \big\langle \ \cZ \doteq  2 M_{+-} + \tfrac12 r \ \big\rangle , \nn\\
\fm_+ &\doteq  \big\langle \ P_+, \ M_{+z} , \ M_{+\zb} , \ K_+  \ \big\rangle ,\qquad
\fm_- \doteq  \big\langle \ P_-, \ M_{-z} , \ M_{-\zb} , \ K_- \ \big\rangle .\nn
\end{align}
Here $r$ is the $\U(1)_r$ generator and $R$, $R^\pm$ are $\SU(2)_R$ generators normalized as in \cite{Cordova_2016} so that $r(Q)=-1$ and $R\in\Z$.%
\footnote{This differs from \cite{Lemos_2020} only for the $\SU(2)_R$ Cartan generator, which is $R\doteq R^1_{\ 1} - R^2_{\ 2}$ for us.}
$\sl_2\oplus\wh\sl_2$ is the algebra of holomorphic plus antiholomorphic conformal transformations of the VOA plane, with the antiholomorphic half twisted by $\SU(2)_R$ rotations.
$\cZ$ generates boosts in the $x^\pm$ plane combined with a $\U(1)_r$ rotation.%
\footnote{Our $\cZ$ is the negative of the $\cZ$ of \cite{Beem_2015, Lemos_2020}.}
Finally, $\fm_\pm$ are nilpotent ideals of the subalgebras in \eqref{Tcohom1},
\begin{align}\label{Tcohom3}
    (\sl_2\oplus\wh\sl_2\oplus\cZ)\circ \fm_\pm 
    \subset \fm_\pm .
\end{align}
The $\fm_\pm$ ideals play a central role in twisted Schur descent.

\subsection{The algebra of the twisted Schur descent operation}\label{sec:2.4}

$\TT_\pm$ descent can be performed with respect to $\TT_\pm$-exact generators with a real action in space-time.  From \eqref{Tcohom2} these are real combinations of $\cZ$ and $\fm_\pm$ for $\TT_\pm$, respectively.
Though the $\wh\sl_2$ generators are $\TT_\pm$-exact, they all have complex action on space-time; real combinations with the $\sl_2$ generators are necessary to generate a real action on space-time.
But since the $\sl_2$ generators are not exact, they are not suitable for performing twisted Schur descent. 
We discuss the effect of real $\sl_2 \oplus \wh\sl_2$ generators on descent operators below.

We now show that, starting from a primary Schur operator $\cO(0)$,%
\footnote{We show at the end of section \ref{sec:2.4} that it is sufficient to start from a primary Schur operator.} applying twisted Schur descent by a sequence of $\TT_+$ and/or $\TT_-$-exact generators produces a series of extended operators which share many of the same properties as their Schur ``parent" $\cO$.  
For unitary SCFTs, Schur operators at the origin satisfy
\begin{align}\label{Schur1}
    \TT_\pm \circ \cO(0) &= 0, &
    &\text{implying}&
    \hL_0\circ\cO(0) & = \cZ\circ\cO(0) =0,
\end{align}
which, from \eqref{Tcohom1}, immediately implies 
\begin{align}
    \wh\sl_2 \circ \cO(0) =_\TT 0.
\end{align} 
Here ``$=_\TT$'' means the equality holds in both $\TT_+$ and $\TT_-$ cohomologies.  
Schur primaries at the origin additionally satisfy
\begin{align}\label{Schur2}
    L_{-1}\circ\cO(0) &= \del_z \cO(0), &
    L_0\circ\cO(0) &= h_\cO \cO(0) ,&
    L_1\circ\cO(0) &= 0,
\end{align}
where the chiral weights are positive half-integers,  $h_\cO\in\N/2$.

Let $\cfE$ denote any operator that satisfies
\begin{align}\label{induction}
    \TT_{\pm} \circ \cfE &= 0, &
    L_{-1} \circ \cfE &=_\TT \del_z \cfE, & 
    L_0 \circ \cfE &=_\TT h_\cfE \cfE , &
    L_1 \circ \cfE &=_\TT 0
\end{align}
from which it also follows $\cZ \circ \cfE = 0$ and $\wh L_0 \circ \cfE =_{\TT} 0$. We will show that twisted Schur descent extended operators $\bD^{N_{\pm}}[\cfE](0)$ automatically satisfy the relations
\begin{align}\label{E1}
   \TT_{\pm} \circ \bD^{N_{\pm}}[\cfE](0) &= 0, &
   \cZ \circ \bD^{N_{\pm}}[\cfE](0) &= 0, \nn\\
   \hL_0 \circ \bD^{N_{\pm}}[\cfE](0) &=_\TT 0, &
   L_0 \circ \bD^{N_{\pm}}[\cfE](0) &=_\TT h_\cfE \bD^{N_{\pm}}[\cfE](0), \\
    L_{-1} \circ \bD^{N_{\pm}}[\cfE](0) &=_\TT \bD^{N_{\pm}}[\del_z \cfE](0), &
    L_1 \circ \bD^{N_{\pm}}[\cfE](0) &=_\TT 0. \nn
\end{align}
These relations inductively imply any twisted Schur descent operators constructed from a Schur operator at the origin, i.e. when $\cfE = \cO(0)$ and $h_{\cfE} = h_{\cO}$, will automatically satisfy the defining relations \eqref{Schur1}, \eqref{Schur2} of a Schur operator in both $\TT_{\pm}$ cohomology.

Let $N_\pm \in \fm_\pm$ be any real vector.  Then, since $N_\pm$ is $\TT_\pm$-exact, there is a fermionic super generator $\n_\pm$ such that 
\begin{align}\label{DA1}
     \TT_\pm \circ \n_\pm = N_\pm .
\end{align}
One then checks that
\begin{align}\label{DA2}
    \TT_\mp \circ \n_\pm &= 0,&
    \TT_\mp \circ N_\pm &= \pm \n_\pm ,
\end{align}
which are related to 
\begin{align}\label{DA3}
    \TT_\pm \circ \TT_\mp &= \cZ, &
    \cZ\circ N_\pm &= \pm N_\pm, &
    \cZ\circ \n_\pm &=\pm \n_\pm,
\end{align}
and that
\begin{align}\label{DA4}
    \n_\pm \circ \n_\pm &= 0,&
    N_\pm \circ \n_\pm &= 0.
\end{align}
Given the definition \eqref{DA1}, the relations \eqref{DA2}, \eqref{DA3}, and \eqref{DA4} follow from the detailed form of the $\cN{=}2$ superconformal algebra, i.e., they do not follow in any obvious way from the automorphisms of that algebra (though half of them follow from the other half by virtue of the anti-linear involution which exchanges $\TT_+ \leftrightarrow \TT_-$). 

Denote by $\bm_\pm$ the fermionic subalgebras satisfying $\TT_\pm\circ\bm_\pm = \fm_\pm$.
These algebras are nilpotent,
\begin{align}\label{DA5}
    \fm_\pm\circ\fm_\pm = \fm_\pm\circ\bm_\pm=\bm_\pm\circ\bm_\pm=0,
\end{align}
(signs correlated), generalizing \eqref{DA4}.
An immediate consequence is that
\begin{align}\label{DA6}
    \bD^{N_\pm} \bD^{N'_\pm} [\cfE] 
    = - \bD^{N'_\pm} \bD^{N_\pm} [\cfE] ,
\end{align}
which follows from the definition of descent \eqref{GD3}.

Assume we have an operator $\cfE$ satisfying \eqref{induction}.
We will show that the most general $\TT_+$ descent operation acting on $\cfE$ produces a new operator, $\bD^X [\cfE]$ as in \eqref{GD3}, which satisfies \eqref{E1}, and therefore also satisfies the induction hypothesis \eqref{induction}.
(The same argument works for $\TT_-$ descent.)
We can choose as the descent generator $X$ any real $\TT_+$-exact generator, which, up to an overall irrelevant rescaling, therefore has one of the three possible forms $\cZ$, $\cZ+N_+$, or $N_+$, for some non-zero $N_+\in\fm_+$.
First,
\begin{align}\label{Zdesc}
    \bD^\cZ[\cfE] 
    &= \int \!\! d\a \,
    \TT_- \circ e^{\a\cZ} \circ \cfE
    = \int \!\! d\a \,
    \TT_- \circ \cfE
    = 0,
\end{align}
where we used \eqref{DA3}, the induction hypotheses \eqref{E1}, and that $\cfE$ is $\TT_-$-closed.
So descent by $\cZ$ gives zero.
Next,
\begin{align}\label{Z+Ndesc}
    \bD^{\cZ+N_+}[\cfE] 
    &= \int_{-\infty}^{+\infty} \!\! d\a \,
    (\TT_-+\n_+) \circ e^{\a(\cZ+N_+)} \circ \cfE
    = \int_{-1}^{+\infty} \!\! dy \,
    \n_+ \circ e^{y N_+} \circ \cfE ,
\end{align}
using that $e^{\a(\cZ+N_+)}=e^{yN_+}e^{\a\cZ}$ with $y=e^\a-1$ which follows from $\cZ\circ N_+=N_+$, and we also used that $e^{-yN_+}\circ\TT_-=\TT_-+y\n_+$ which follows from $\TT_-\circ N_+=\n_+$ and $\n_+\circ N_+=0$.
The final expression in \eqref{Z+Ndesc} is like $N_+$ descent except with a boundary at finite points in space-time.
We therefore expect it not to be $\TT_+$-closed.
Indeed, by the same argument as in \eqref{GD4}, $\TT_+ \circ \bD^{\cZ+N_+}[\cfE] = e^{yN_+}\circ\cfE |^{y=\infty}_{y=-1} \neq 0$.  So descent by $\cZ+N_+$ fails to give an operator in cohomology.

We thus only ever need to consider descent operators of the form
\begin{align}\label{Ndesc1}
    \bD^{N_+}[\cfE] 
    &= \int \!\! d\a \,
    \n_+ \circ e^{\a N_+} \circ \cfE ,&
    N_+ &\in \fm_+ .
\end{align}
A similar formula defines $\bD^{N_-}[\cfE]$ for $N_-\in\fm_-$.

We now show $\bD^{N_+}[\cfE]$ is in both $\TT_\pm$ cohomologies.
It is in $\TT_+$-cohomology by the generalized descent construction up to boundary terms as in \eqref{GD4}.
These boundary terms need to be evaluated for particular choices of $N_+$ and $\cfE$.
Assume for the moment that these boundary terms vanish.
Then we compute
\begin{align}\label{Ndesc2}
    \TT_- \circ \bD^{N_+}[\cfE]
    &= - \int_{-\infty}^{+\infty} \!\! d\a \,
    \n_+ \circ e^{\a N_+} \circ (\TT_-+\a \n_+) \circ\cfE 
    = 0 ,
\end{align}
where in the first step we used the descent algebra identities \eqref{DA1} and \eqref{DA2} to commute $\TT_-$ to the right, and the vanishing follows from the induction hypothesis that $\cfE$ is $\TT_-$-closed and from the nilpotency relations \eqref{DA4}.
The identity \eqref{Ndesc2} --- which does not have an analog in topological descent --- is crucial in allowing twisted Schur descent to extend the vertex operator algebra of twist-translated Schur operators to the whole descent web. 
Unlike in the case of Schur operators, which are in both $\TT_+$ and $\TT_-$ cohomology by virtue of unitarity, this property of the descent operators instead follows in a less obvious way by virtue of the detailed structure of the superconformal algebra.

Note that the $\U(1)_r$ charges of the twisted Schur supercharges are $r(\TT_\pm)=\mp1$, so by \eqref{DA1}, $r(\n_\pm)=\pm1$ since bosonic generators have zero $r$-charge.
Then it follows from \eqref{Ndesc1} that
\begin{align}\label{r-charge}
    r(\bD^{N_\pm}[\cfE]) = r(\cfE)\pm1 .
\end{align}

Note also that, using the relation $\n_+ = \TT_-\circ N_+$ in the definition of $\bD^{N_+}[\cfE]$ and commuting a $\TT_-$ to the right to annihilate $\cfE$, one derives an identity of the form $\bD^{N_+}[\cfE] = \bD^{N_+}[\cfE] +$boundary terms.
So these boundary terms must vanish if $\bD^{N_+}[\cfE]$ is well-defined, giving
\begin{align}\label{Ndesc3}
    \a \n_+ \circ e^{\a N_+} \circ \cfE \big|^{\a=+\infty}_{\a=-\infty}
    &= \TT_- \circ \left( e^{\a N_+}\circ \cfE \big|^{\a=+\infty}_{\a=-\infty} \right) .
\end{align}
Since the boundary terms on the right vanish by \eqref{GD4} for $\bD^{N_+}[\cfE]$ to be $\TT_+$-closed, it follows that the apparently less convergent boundary terms on the left must also vanish.

We now show that $\bD^{N_+}[\cfE]$ satisfies the induction hypotheses \eqref{induction} if $\cfE$ does.
It immediately follows that $\cZ\circ \bD^{N_+}[\cfE] = \wh\sl_2\circ \bD^{N_+}[\cfE] =_\TT 0$ since $\cZ$ and $\wh\sl_2$ are both $\TT_\pm$-exact.
To show the $L_0\circ\bD^{N_+}[\cfE]$ and $L_1\circ\bD^{N_+}[\cfE]$ relations we need the action of $\sl_2$ on the $\fm_\pm$ and $\bm_\pm$ subalgebras.
Because $\sl_2$ is $\TT_\pm$-closed, if $N_\pm\in\fm_\pm$ and $\n_\pm\in\bm_\pm$ are related by $\TT_\pm \circ \n_\pm = N_\pm$, then for any element $X\in\sl_2$, $N'_\pm \doteq X\circ N_\pm$ and $\n'_\pm \doteq X\circ \n_\pm$ are related in the same way.

Consider $X\in \sl_2$.  By commuting $X$ to the right one finds%
\footnote{We are violating our associativity convention in writing the last term which is meant to be $N'_+\circ \n_+\circ e^{\a N_+} \circ \cfE$, and not $(N'_+\circ \n_+)\circ e^{\a N_+} \circ \cfE$ which vanishes identically.}
\begin{align}\label{L0E}
    X \circ \bD^{N_+}[\cfE] 
    &=_\TT \bD^{N_+} [X\circ\cfE]
    + \int \!\! d\a (\n'_+ {+} \a N'_+{\circ}\n_+) {\circ} e^{\a N_+} {\circ} \cfE  .
\end{align}
Using $\n'_+=\TT_-\circ N'_+$, the integrated term in \eqref{L0E} is 
\begin{align}
    \int \!\! d\a \cdots
    &=_{\TT_-} \int \!\! d\a N'_+{\circ} (-\TT_- {+} \a \n_+) {\circ} e^{\a N_+} {\circ} \cfE 
    = -\int \!\! d\a N'_+{\circ} e^{\a N_+} {\circ} (\TT_- {+} \a \n_+ {-}\a \n_+) {\circ}  \cfE \nn
\end{align}
which vanishes in $\TT_-$-cohomology.
Similarly, using $N'_+=\TT_+\circ \n'_+$, $N_+=\TT_+\circ \n_+$, and $\TT_+\circ e^{\a N_+}\circ\cfE=0$, this term is, in $\TT_+$-cohomology, the boundary term
\begin{align}
    \int \!\! d\a \cdots
    &=_{\TT_+} \int \!\! d\a \n'_+{\circ} (1 {+} \a N_+) {\circ} e^{\a N_+} {\circ} \cfE 
   = \int \!\! d\a \frac{d}{d\a} \left\{\a \n'_+{\circ} e^{\a N_+} {\circ} \cfE \right\} 
   = \a \n'_+ {\circ} e^{\a N_+} {\circ} \cfE \Big|^{+\infty}_{-\infty}
   .\nn
\end{align}
This boundary term is closely related to the boundary term in \eqref{Ndesc3} that vanishes for $\bD^{N_+}[\cfE]$ to be a well-defined operator in $\TT_\pm$-cohomology, making it plausible that it, too, vanishes.
But we do not have a general argument proving this, and resort to checking it on a case-by-case basis. 
For all the descent operators we construct for the free hyper multiplet in section \ref{sec:4}, we find these boundary terms vanish, and so obey
\begin{align}\label{L0E1}
    X \circ \bD^{N_+}[\cfE] 
    &=_\TT \bD^{N_+} [X\circ\cfE] .
\end{align}
We will assume this is true in general from now on.
The upshot is that, by taking $X\in\{L_{-1},L_0,L_1\}$ in \eqref{L0E1} and assuming the induction hypothesis \eqref{induction}, we have shown the $\bD^{N_+}[\cfE]$ descent operator satisfies \eqref{E1}. 

Note that when we take $\cfE = \cO(0)$ to be a Schur primary and let $X = L_{-1}$, \eqref{L0E1} inductively implies that any descent operator constructed from a Schur descendant $(L_{-1})^n\circ \cO(0)$ is equivalent to $(L_{-1})^n$ acting on the same descent operator constructed from the Schur primary $\cO(0)$. Thus, it is sufficient to only consider descent operators constructed from Schur primaries.


\subsection{Cohomological equivalence among descent operators}\label{sec:2.5}

We have now reduced each step of the twisted Schur descent procedure \eqref{Ndesc1} to choosing an element $N_+\in\fm_+$ or an $N_-\in\fm_-$ with which to perform descent.  
Since $\fm_\pm$ are 4-dimensional vector spaces, there is an $\R\P^3$ of inequivalent choices of $N_\pm$, since the overall scale of $N_\pm$ can be absorbed into the integration measure of the descent operation.
We will now show that almost all these choices are, in fact, equivalent in $\TT_\pm$-cohomology, and only a finite set of choices of $N_\pm\in\fm_\pm$ needs to be considered.

The key is to consider the action of the $\wh\SO(1,3)$ group of twisted conformal transformations of the VOA plane on descent operators.  $\wh\SO(1,3)$ is generated by elements of $\sl_2\oplus\wh\sl_2$ with real space-time action.  ``Twisted" refers to the accompanying $\SU(2)_R$ transformation that is part of $\wh\sl_2$.
To this end, define the superconformal group elements
\begin{align}\label{ghrels1}
    \hg(z) &\doteq \exp\{ z L_{-1} + \zb \hL_{-1} \}, &
    &\text{twist translations},\nn\\
    \hg(\ell) &\doteq \exp\{ \ell L_0 + \ellb \hL_0 \}, &
    &\text{twist rotation and dilatation},\\
    \hg(\b) &\doteq \exp\{ \b L_1 + \bb \hL_1 \}, &
    &\text{twist special conformal transformations},\nn
\end{align}
for $z$, $\ell$, and $\b$ arbitrary complex numbers, which act on the VOA plane as indicated above. 
Consider, say, $\hg(\b) \circ \cfE = e^{\b L_1}e^{\bb \hL_1}\circ\cfE$ where $\cfE$ satisfies the induction hypothesis \eqref{induction}.  
Since $\cfE$ is $\TT_\pm$-closed and $\hL_1$ is $\TT_\pm$-exact, we have $\hg(\b) \circ \cfE =_\TT e^{\b L_1}\circ\cfE$, by which we simply mean that the cohomology class of $\hg(\b) \circ \cfE$ depends only on $\b$.%
\footnote{This does not mean that the $e^{\bb \hL_1}$ part of the twist special conformal transformation can be neglected: it is needed to create a physical operator in Minkowski space. This holds for all $\hg$ twist transformations.}
Furthermore, since $L_1\circ\cfE=_\TT0$ by assumption, we have $\hg(\b) \circ \cfE =_\TT \cfE$.  

A similar argument for $\hg(\ell)$ and $\hg(z)$ using \eqref{induction} implies
\begin{align}\label{ghrels2}
    \hg(\b)\circ\cfE &=_\TT \cfE, &
    \hg(\ell)\circ\cfE &=_\TT e^{\ell h_\cfE} \cfE, &
    \hg(z)\circ\cfE &=_\TT e^{zP_z}\circ\cfE.
\end{align}
The $\hg(z)$ relation shows, just as for Schur operators, that the twisted Schur cohomology class of $\cfE$ only depends on the holomorphic $z$ coordinate of the VOA plane.
The $\hg(\ell)$ and $\hg(\b)$ relations are much stronger, implying essentially that $\cfE$ cohomology classes are independent of both $\ell$ and $\ellb$ as well as $\b$ and $\bb$.

Applying the twisted $\hg$ transformations to $\bD^{N_+}[\cfE]$ gives
\begin{align}\label{ghrels2.5}
\hg(\b) \circ \bD^{N_+}[\cfE] & =_{\TT} \bD^{N_+}[\cfE] \nn \\
\hg(\ell) \circ \bD^{N_+}[\cfE] &=_{\TT} e^{\ell h_{\cfE}}\bD^{N_+}[\cfE] \\
\hg(z) \circ \bD^{N_+}[\cfE] &=_{\TT} \bD^{N_+}[e^{z P_z}\circ \cfE] \nn
\end{align}
where we used \eqref{L0E1} to commute the $\hg$'s past the descent operation in $\TT_\pm$ cohomology, and the relations in \eqref{ghrels2} for $\cfE$ and \eqref{GD5}. 

Since $\fm_\pm$ are ideals of the superconformal algebra cohomology subalgebras, \eqref{Tcohom3}, it follows that
\begin{align}\label{ghrels3}
    \hg \circ \bD^{N_+} [\cfE] 
    &=\int \!\! d\a \,(\hg\circ\n_+) \circ e^{\a \, \hg\circ N_+} \circ (\hg\circ\cfE)
    = \bD^{\hg\circ N_+} [\hg\circ\cfE].
\end{align}
Then, using \eqref{GD5} and the relations in \eqref{ghrels2} gives
\begin{align}\label{ghrels3.5}
\hg(\b) \circ \bD^{N_+}[\cfE] & =_{\TT} \bD^{\hg(\b)\circ N_+}[\cfE] ,\nn \\
\hg(\ell) \circ \bD^{N_+}[\cfE] &=_{\TT} e^{\ell h_{\cfE}}\bD^{\hg(\ell) \circ N_+}[\cfE] ,\\
\hg(z) \circ \bD^{N_+}[\cfE] &=_{\TT} \bD^{\hg(z) \circ N_+}[e^{z P_z}\circ \cfE] .\nn
\end{align}
Combining \eqref{ghrels3.5} and \eqref{ghrels2.5} implies
\begin{align}\label{ghrels4}
    \bD^{N_+} [\cfE] 
    &=_\TT \bD^{\hg(\b)\circ N_+} [\cfE], \nn\\
    \bD^{N_+} [\cfE] 
    &=_\TT \bD^{\hg(\ell)\circ N_+} [\cfE], \\
    \bD^{N_+} [e^{zP_z} \circ \cfE] 
    &=_\TT \bD^{\hg(z)\circ N_+} [e^{zP_z} \circ\cfE]. \nn
\end{align}
Note that the last equation in \eqref{ghrels2.5} shows that twist-translation commutes with descent, while the last equation in \eqref{ghrels4} relates descent operators constructed from twist-translations of cohomology classes $\cfE$.

Now compute the $\hg(z),\hg(\ell),\hg(\b)$ actions on $\fm_\pm$ explicitly.
Parameterize $N_+$ as
\begin{align}\label{ghrels5}
    \fm_+ \ni N_+ 
    &\doteq p P_+ + m M_{+z} + \bar m M_{+\zb} + k K_+,&
    p , k &\in \R, &
    m &\in \C.
\end{align}
A short calculation gives
\begin{align}\label{ghrels6}
    \hg(\b): \big( p,m,k \big) &\mapsto 
    \big(\ p \, , \, 
    m {+} 2p\bb \, , \,
    k {-} \Re(m\b) {-} p \b\bb \ \big) ,
    \nn\\
    \hg(\ell): \big( p,m,k \big) &\mapsto 
    e^{(\ell+\ellb)/2} \,\big( \ p \, , \, 
    e^{-\ellb} m \, , \,
    e^{-\ell-\ellb}k \ \big) ,
    \\
    \hg(z): \big( p,m,k \big) &\mapsto 
    \big(\ p {-} \Re(m\zb) {-} kz\zb \, , \, 
    m {+} 2kz \, , \,
    k \ \big) .
    \nn
\end{align}
One can show that if $p\neq0$ there is a sequence of $\hg(\b)$ and $\hg(\ell)$ transformations which takes $(p,m,k)\to (1,0,\sgn k)$.  Likewise, if $p=0$ and $m\neq0$ one can set $(p,m,k)\to(0,1,0)$, and if $p=m=0$ one can set $(p,m,k)\to(0,0,1)$.
There is a similar result for $\fm_-$.
Thus, at any stage there are just 10 possible cohomologically inequivalent descent operations, 
\begin{align}\label{ghrels7}
    \bD^{P_\pm}, \bD^{M_{\pm3}}, \bD^{K_\pm}, \bD^{P_\pm+K_\pm}, \qq{and} \bD^{P_\pm-K_\pm}.
\end{align}

There is a subtle issue regarding the meaning of equivalence between descent operators. What we have just shown is the following: assuming the induction hypothesis \eqref{induction}, if one considers two descent operators $\bD^{X_1}[\cfE]$, $\bD^{X_2}[\cfE]$ constructed using two distinct descent operations from \eqref{ghrels7}, then there is no inner automorphism of the superconformal algebra which maps $\bD^{X_1}[\cfE] =_{\TT} \bD^{X_2}[\cfE]$ in cohomology. However, this result does not imply $\bD^{X_1}[\cfE]$ and $\bD^{X_2}[\cfE]$ are automatically inequivalent in cohomology. This is because additional equivalences can exist that come in two types:
\begin{outline}
\1[] \textbf{4d-type equivalence} $=$ equivalences between 4d operator expressions in the 4d twisted Schur cohomology theory (which could be equalities or only hold in cohomology);
\1[] \textbf{VA-type equivalence} $=$ equivalences between twisted Schur cohomology classes at the level of OPEs and correlation functions of the vertex algebra.
\end{outline}
4d-type equivalences could occur due to simplifications that result from additional properties the parent operator $\cfE$ possesses that go beyond those assumed in \eqref{induction}. 
We will see instances of both additional equalities and additional cohomology equivalences between operator expressions in the next section. VA-type equivalences, on the other hand, occur when two descent operators have identical OPEs and correlation functions within the vertex algebra.
A VA-type equivalence between two operators is the same as saying that their difference is a null state in the vertex algebra.
A 4d-type equivalence between two operators implies a VA-type equivalence between them while the converse is not true.

Without detailed structural knowledge of the vertex algebra, such as the operators defining a generating set for it, the question of a VA-type equivalence is difficult to prove a priori from 4d. 
A VA-type equivalence can also appear between a descent operator and a local Schur operator. 
In fact, such an equivalence already emerges in the $\cO$-$\bL^\pm$-$\bS$ subalgebra for the free hypermultiplet we present in section \ref{sec:4}. 

As we derive the descent web of figure \ref{web} in the next section, we will be careful to identify where a 4d-type or VA-type equivalence might remain between various descent operators.


\subsection{The descent web}\label{descent_web}

We now map out the possible inequivalent descent operators which constitute figure \ref{web}.

To begin, we show that a necessary condition for a descent operator $\bD^X[\cfE]$ to be well-defined as an extended operator is that $X$ must translate the world volume of the parent operator $\cfE$ as a set.%
\footnote{The action of the generator $X \in \fm_{\pm}$ may still have fixed points on the $\cfE$ world volume, as is the case for $X=M_{\pm3}$.} 
By \eqref{Ndesc1}, the $\bD^{K_\pm}[\cO]$ and $\bD^{M_{\pm3}}[\cO]$ descents of a Schur operator at the origin are $\int_{-\infty}^{+\infty} d\a$ integrals of some local operators at the origin, because $e^{\a K_\pm}$ and $e^{\a M_{\pm3}}$ do not move the origin.
Since $K_\pm\circ\cO=0$ for $\cO$ primary, the local operator is just $\k_\pm\circ \cO$, where $\k_\pm$ are the descent supercharges $\n_\pm$ associated to $K_\pm$.
So $\bD^{K_+}[\cO] = (\infty)\cdot\k_\pm\circ \cO$.
The operator $e^{\a M_{\pm3}}\circ\cO$ is an $\a$-weighted linear combination of spin components of the Lorentz multiplet that $\cO$ belongs to, so it is at least as divergent for a scalar Schur operator and more divergent for Schur operators with spins $j>0$ or $\tj>0$. 
The conclusion is that these descent procedures do not produce convergent extended operators.
A similar pattern applies to each successive descent operation $\bD^X[\cfE]$:  if $X$ does not translate the world volume of $\cfE$ then the descent diverges, and if it does, it potentially generates a convergent extended operator of one greater world volume dimension.

Starting from a Schur operator at the origin, we now map out the possible line, surface, and domain wall world volumes that can be obtained by successively combining the ten inequivalent descent operations of \eqref{ghrels7}.

\paragraph{Lines.} 

The transformations $e^{\a P_\pm}$ acting on the origin generate the lightlike lines
\begin{align}\label{dw2}
    L^\pm \doteq \{ x^\mp=z=\zb=0 \}.
\end{align}
Also, $e^{\a (P_+ \pm K_+)}$ generate the $L^+$ line while $e^{\a(P_- \pm K_-)}$ generate the $L^-$ line.
The remaining generators, $M_{\pm3}$ and  $K_\pm$, do not move the origin, so there are no further 1d world volumes.

\paragraph{Surfaces.} 

$e^{\a P_\mp}$ acting on $L^\pm$ (signs correlated) generate the plane,
\begin{align}\label{dw3}
    S \doteq \{ z=\zb=0 \},
\end{align}
which can also be generated by $e^{\a(P_+ \pm K_+)}$ acting on $L^-$ and $e^{\a(P_- \pm K_-)}$ acting on $L^+$.
Acting with $e^{\a M_{\mp3}}$ on $L^\pm$ (signs correlated) both generate the cone,
\begin{align}\label{dw4}
    \hS \doteq \{ x^+x^-+z\zb=0, \, z=\zb \},
\end{align}
which is the $z=$ real ($x^4=0$) slice of the light cone in Minkowski space-time.
The remaining generators, $P_\pm$, $M_{\pm3}$, $K_\pm$, and $K_\mp$, do not move $L^\pm$, so there are no further 2d world volumes.

\paragraph{Domain walls.} 

The action of $e^{\a P_\pm}$, $e^{\a(P_+ \pm K_+)}$, $e^{\a(P_- \pm K_-)}$ on $\hS$ and $e^{\a M_{\pm3}}$ acting on $S$ all generate the domain wall (3d hyperplane)
\begin{align}\label{dw5}
    \wh W \doteq \{ z=\zb \}.
\end{align}
The remaining generators do not move $S$ or $\hS$.

\paragraph{4d volume.} 

None of the generators move the $\wh W$ world volume, so the descent procedure terminates after the third order and we obtain no volume operators.\\

\begin{figure}[ht]
\centering
\begin{tikzpicture}[>=stealth']
\draw[->,thick,draw=black] (0,-2.9) -- (0,3.4);
\draw[->,rotate=-90,thick,draw=black!40] (0,-2.9) -- (0,2.9);
\draw[->,rotate=-60,thick,draw=black] (0,-2.7) -- (0,2.7);
\draw[thick,draw=violet,fill=violet!15,fill opacity=0.3] (-2,1.6) -- (2,3.5) -- (2,-1.5) -- (-2,-3.6) -- cycle;
\draw[thick,draw=red!35,fill=red!10,fill opacity=0.3] (0,0) -- (2.5,-2.5) arc (375:165:2.5cm and .5cm) -- cycle;
\draw[thick,dashed,red!35] (2.5,-2.5) arc (15:165:2.5cm and .5cm);
\draw[rotate=180,thick,draw=red!35,fill=red!10,fill opacity=0.3] (0,0) -- (2.5,-2.5) arc (375:165:2.5cm and .5cm) -- cycle;
\draw[rotate=180,thick,red!45] (2.5,-2.5) arc (15:165:2.5cm and .5cm);
\draw[rotate=110,ultra thick,draw=blue] (0,0) -- (2.3,0);
\draw[rotate=-70,ultra thick,draw=blue!50] (0,0) -- (2.3,0);
\draw[rotate=-110,ultra thick,draw=blue] (0,0) -- (3.25,0);
\draw[rotate=70,ultra thick,draw=blue!50] (0,0) -- (3.25,0);
\node[circle,fill=green!70!black,scale=.7] at (0,0) {};
\draw[->,thick,draw=black] (0,2.1) -- (0,3.4);
\draw[->,rotate=-90,thick,draw=black] (0,0.1) -- (0,2.9);
\draw[rotate=-90,thick,draw=black] (0,-2.9) -- (0,-2.0);
\node at (0,3.6) {$x^1$};
\node at (2.6,1.4) {$x^2$};
\node at (4.5,0) {$x^3\subset$ VOA plane};
\node[green!70!black] at (-0.5,0) {$\cO$};
\node[blue] at (-0.8,1.3) {$L^-$};
\node[blue] at (-1.0,-1.8) {$L^+$};
\node[violet] at (1.5,0.4) {$S$};
\node[red] at (-1.5,2.7) {$\hS$};
\end{tikzpicture}
\caption{The possible world volumes of twisted Schur descent operators.  
The figure shows the $x^4=0$ 3d slice of Minkowski space-time; $x^1$ is the time coordinate.
The $x^3$ axis is the $z=\zb$ line in the VOA plane.
$L^\pm$ and $\hS$ are in the light cone.
The $\wh W$ 3d world volume fills the 3d slice shown in the figure.
The extension of the $\hS$ and $\wh W$ world volumes in the $x^3$-direction of the VOA plane is conventional: this direction can be rotated while keeping the corresponding descent operators in the same cohomology class.
}
\label{fig1}
\end{figure}
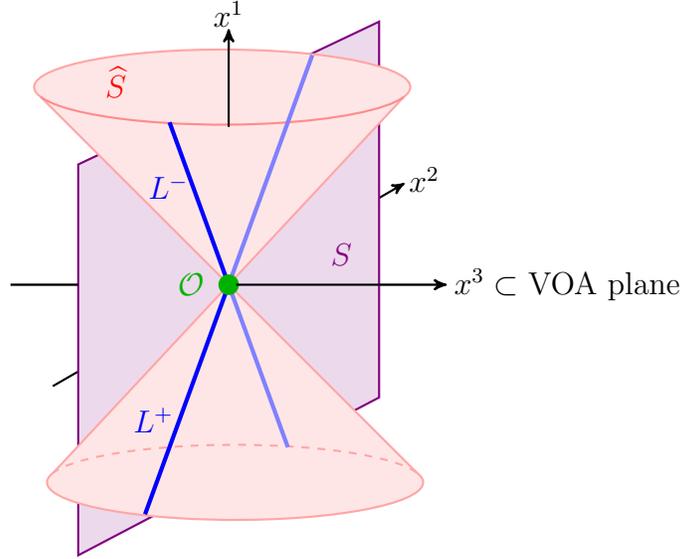

\noindent These world volumes are illustrated in figure \ref{fig1}, which shows the $x^4=0$ ($z=\zb$) slice of Minkowski space-time.

A descent operator is not defined solely by its  world volume just as a local operator is not solely defined by its location in space-time. 
Hence, despite there being multiple routes to producing a given world volume, it is possible the associated descent operators are inequivalent as operators.
Starting with the descent line operators, we now describe the inequivalent descent operators that persist among the various routes one can take to obtain a given world volume.
For each inequivalent descent operator, we will denote the result by a corresponding boldfaced symbol, e.g. $\bL^{\pm}[\cO]\doteq \bD^{P_{\pm}}[\cO]$ or $\bS[\cO] \doteq \bD^{P_-} \bD^{P_+}[\cO]$.

\paragraph{Lines.}

Since $P_\pm$ translate the origin along the $x^\pm$ axes, the $\bD^{P_\pm}[\cO]$ descent operators are light-like line operators.
We discuss the convergence of these line operators in more detail in the next section.
The descent operator $\bD^{P_+ \pm K_+}[\cO]$ is equivalent to $\bD^{P_+}[\cO]$, which we can see through a brief computation:
\begin{align}\label{dw1}
    \bD^{P_+\pm K_+}[\cO] 
    &= \int \!\! d\a (\pi_+\pm\k_+) \circ e^{\a(P_+\pm K_+)} \circ \cO
    = \int \!\! d\a (\pi_+\pm\k_+) \circ e^{\a P_+} \circ \cO
    \nn\\
    &= \int \!\! d\a e^{\a P_+} \circ (\pi_+\pm\k_+) \circ \cO
    = \int \!\! d\a e^{\a P_+} \circ \pi_+ \circ \cO
    = \bD^{P_+}[\cO].
\end{align}
Here, $\TT_+\circ\pi_+ = P_+$, $\TT_+\circ\k_+ = K_+$, and we have used the fact that $K_+\circ \cO=0$ for a Schur primary in the second equality and the nilpotency relations \eqref{DA5} in the third equality. 
The final result follows from the fact $\k_+\circ \cO=0$ for a Schur primary, which is explained by the relation $\k_+ = \TT_-\circ K_+$ from \eqref{DA2}, which implies $\k_+\circ \cO = (\TT_-\circ K_+) \circ \cO = \TT_-\circ K_+ \circ \cO - K_+ \circ \TT_-\circ \cO = 0$ since a Schur primary is annihilated by $K_+$ and by $\TT_-$.
A similar computation indicates $\bD^{P_-\pm K_-}$ descent is equivalent to $\bD^{P_-}$ descent.
Thus, at the first level of the descent web, the only inequivalent, convergent descent operators are the two lightlike line operators
\begin{align}\label{L+-descent}
    \bL^+[\cO] \doteq \bD^{P_+}[\cO] \qand \bL^-[\cO] \doteq \bD^{P_-}[\cO].
\end{align} 
Furthermore, we know these two operators must be inequivalent at the level of the vertex algebra because, from \eqref{r-charge}, they necessarily have opposite $U(1)_r$ charges, as indicated by their superscript.

\paragraph{Surfaces.} 

As indicated in the descent web of figure \ref{web}, the descent operations $\bD^{P_{\pm}} \bL^{\mp}[\cO]$ give equivalent descent operators. 
This follows from the anti-commutativity of the $\bD^{P_+}$ and $\bD^{P_-}$ descent operations. 
To see this, let $\pi_\pm$ be the fermionic supercharges satisfying $\TT_\pm\circ\pi_\pm=P_\pm$ in the $\bD^{P_{\pm}}$ descent procedures.  
Then by \eqref{DA2}, $\pi_\pm=\pm\TT_\mp\circ P_\pm$.
Substituting these in $\pi_+\circ\pi_-$, using the Jacobi identities, that $\TT_\pm\circ P_\pm=0$, and that $P_+\circ P_-=0$, one finds that $\pi_+\circ\pi_- = - P_+\circ (\cZ \circ P_-) = P_+\circ P_- =0$ using \eqref{DA3} in the last steps.
Then $\bD^{P_+}\bD^{P_-}[\cO]=-\bD^{P_-}\bD^{P_+}[\cO]$ follows immediately from the definition \eqref{Ndesc1} of the descent operation.
We denote the resulting surface operator by
\begin{align}\label{Sdescent}
    \bS[\cO] \doteq \bD^{P_-} \bD^{P_+}[\cO] = \bD^{P_-}\bLp[\cO]. 
\end{align}
In \eqref{dw1} we showed that $\bD^{P_+\pm K_+}[\cO] = \bD^{P_+}[\cO]$, but
this relation is much harder to show at higher levels of descent. 
Nonetheless, for surfaces it is possible to show the equivalences
\begin{align}\label{S_equivs}
    \bD^{P_-\pm K_-}\bLp[\cO] =_\TT \bS[\cO] \qand \bD^{P_+\pm K_+}\bLm[\cO] =_\TT -\bS[\cO].
\end{align}
These results follow from a lengthy calculation and hold only in cohomology; the details will be reported in \cite{Argyres_2023xx}.

Next, denote the descent operators $\bD^{M_{+3}}\bLm[\cO]$ and $\bD^{M_{-3}}\bLp[\cO]$ by
\begin{align}\label{hbSLR}
 \hbS_L[\cO] \deq \bD^{M_{+3}}\bLm[\cO], \qquad  \hbS_{R}[\cO] \deq \bD^{M_{-3}}\bLp[\cO].
\end{align}
The $\hbS_{\ldots}$ descent operators are not equivalent to the $\bS$ descent operator. Intuitively, this makes sense since they don't have equal world volumes, but sufficient evidence stems from the fact that their correlators display different behaviors at the level of the vertex algebra. Specifically, $\hbS_{\ldots}$ correlators always come with a quasi-topological ``cut", while those of $\bS$ do not.

Furthermore, it can be proven that $\hbS_L$ and $\hbS_R$ are inequivalent cohomology classes as operator expressions. 
This is an intricate result which will be discussed in \cite{Argyres_2023xx}, but it is a direct consequence of the regularization prescription needed to define hatted operators in general. 
We address the convergence properties and required regularization for the hatted operators in section \ref{sec:3}. 

Lastly, although inequivalent as  operators, preliminary evidence suggests that the VA-type equivalence $\hbS_L[\cO] = - \hbS_R[\cO]$ may still be consistent at the level of the vertex algebra. 
We explain the consequences of this scenario for the descent web in the next section, and will report on its viability more thoroughly in \cite{Argyres_2023xx}.

\paragraph{Domain walls.} 

Among the ten possible routes to producing the $\wh W$ world volume from the $S$ and $\wh S$ surfaces, the two equivalences
\begin{align}
\bD^{P_+} \hbS_L[\cO] = - \bD^{M_{+3}}\bS[\cO] \qand \bD^{P_-} \hbS_R[\cO] = - \bD^{M_{-3}}\bS[\cO]
\end{align}
follow directly from the nilpotency relation \eqref{DA6}. 
The remaining eight routes produce apparently inequivalent extended operators which we denote by
\begin{align}
    \hbW_L^-[\cO] &\deq \bD^{P_-}\hbS_L[\cO], \quad &\hbW_{L\pm}^-[\cO] &\deq \bD^{P_- \pm K_-}\hbS_{L}[\cO], \quad &\hbW_L^+[\cO] &\deq \bD^{P_+}\hbS_{L}[\cO], \nn \\
    \hbW_R^+[\cO] &\deq \bD^{P_+}\hbS_R[\cO], \quad &\hbW_{R\pm}^+[\cO] &\deq \bD^{P_+ \pm K_+}\hbS_{R}[\cO], \quad &\hbW_R^-[\cO] &\deq \bD^{P_-}\hbS_{R}[\cO].
\end{align}
Depending on the VA-type equivalence between the $\hbS_L$ and $\hbS_R$ surface operators, there can be a number of additional equivalences between the wall operators that would result in only two distinct equivalence classes which differ in their $U(1)_r$ charge. Apart from this, there could also be additional 4d-type equivalences among certain domain wall operators. We elaborate on these possibilities in the next section.

\subsection{Possible equivalences and their consequences}

For the $\hbS_L$ and $\hbS_R$ descent operators, we have evidence in form of vertex algebra OPEs and 2-point functions (for the free hypermultiplet and free vector multiplet) which relate them by the VA-type equivalence, 
\begin{align}\label{SL=-SR}
    \hbS_L[\cO] \overset{\text{VA-type}}{\equiv} - \hbS_R[\cO] .
\end{align}
Assuming this equivalence holds for general Schur operators, it implies the following equivalences among the domain wall operators:
\begin{align}
\hbW^-_{L\pm}[\cO] &=_{\TT} \hbW^-_R[\cO],  &\hbW_L^-[\cO] &=  \hbW_R^-[\cO] , \nn \\
\hbW^+_{R\pm}[\cO] &=_{\TT} - \hbW^+_L[\cO],  &\hbW_R^+[\cO] &= - \hbW_L^+[\cO]. 
\end{align}
We can prove the first line of equivalences all at once by considering the descent operator $\bD^{P_- + \b K_-}\hbS_L$ for $\b \in \{ 0, \pm 1\}$ which coincides with $\hbW_{L\pm}^-$ when $\b = \pm 1$ and $\hbW_L^-$ when $\b = 0$. 
We have,
\begin{align}
 \bD^{P_- + \b K_-} \hbS_L &\overset{\text{VA-type}}{\equiv} -\bD^{P_- + \b K_-}\hbS_R = - \bD^{P_- + \b K_-}\bD^{M_{-3}} \bLp \nn \\
& \quad = +\bD^{M_{-3}}\bD^{P_- + \b K_-} \bLp =_{\TT} \bD^{M_{-3}}\bS = \hbW^-_R.
\end{align}
The 2nd equality follows from the definition of $\hbS_R$, the 3rd equality follows from the nilpotency relation \eqref{DA6}, and the 4th equality only holds in cohomology for $\b = \pm 1$ after using the equivalence \eqref{S_equivs} between $\bD^{P_-\pm K_-} \bD^{P_+}$ and $\bS$. Using $\bD^{P_+ + \b K_+} \hbS_R$ to denote $\hbW^+_{R\pm}$ when $\b = \pm 1$ and $\hbW^+_R$ when $\b = 0$, an identical computation can be done to prove the other two domain wall equivalences. 

Thus, with the assumption of the VA-type equivalence \eqref{SL=-SR}, we learn that there exist only two inequivalent domain wall descent operators: $\hbW_R^-[\cO]$ and $\hbW_L^+[\cO]$. We can be certain these two operators are inequivalent at the level of the vertex algebra because they have opposite $\U(1)_r$ charges.

In the event that the equivalence \eqref{SL=-SR} is not true, it is still possible for there to be 4d-type equivalences among certain domain walls as well as VA-type equivalences. A necessary condition for either equivalence type to exist is the corresponding wall operators must have the same $U(1)_r$ charge. Given the equivalence between $\bD^{P_- \pm K_-} =_{\TT}\bD^{P_-}$ for Schur operators \eqref{dw1} and descent lines \eqref{S_equivs}, the most likely domain wall scenario is to have the analogous 4d-type equivalences,
\begin{align}
    \hbW^-_{L\pm}[\cO] =_{\TT} \hbW^-_L[\cO] \qand \hbW^+_{R\pm}[\cO] =_{\TT} \hbW^+_R[\cO].
\end{align}
It is possible (though unlikely) to have an additional 4d-type equivalence between $\hbW_L^-$ and $\hbW^-_R$, though we have not checked this.


\subsection{VOA plane twist translations and special conformal transformations}

So far we have performed descent operations on Schur primaries at the origin of the VOA plane.
These are twist-translated to other points on the VOA plane using the $\hg(z)$ twist translations of \eqref{ghrels1}. We denote the resulting descent operator by
\begin{align}
    \bD^X[\cfE](z) \deq \hg(z) \circ \bD^X[\cfE](0),
\end{align}
where it should be understood that $\bD^X[\cfE](0) \equiv \bD^X[\cfE]$.
For example, 
\begin{align}
\bLp[\cO](z) = \exp(z L_{-1} + \zb \wh L_{-1}) \circ \bLp[\cO](0).
\end{align}
By virtue of the last relation in \eqref{ghrels2.5}, twist-translations commute with all descent operations. 
Thus, we can apply this relation iteratively to a twist-translated descent operator to always rewrite it as the same descent operator being constructed from a twist-translated Schur operator. 
It follows that, 
\begin{align}
    \bL^\pm[\cO](z) &=_{\TT} \bL^\pm[\cO(z)], &
    \bS[\cO](z) &=_{\TT} \bS[\cO(z)], \\ \nn
    \hbS_{\ldots}[\cO](z) &=_{\TT} \hbS_{\ldots}[\cO(z)] , & \hbW^{\pm}_{\ldots}[\cO](z) &=_{\TT} \hbW^{\pm}_{\ldots}[\cO(z)].
\end{align}
where it is also understood that a twist-translated Schur operator is denoted as,
\begin{align}
    \cO(z) \deq \exp(z L_{-1} + \zb \wh L_{-1})\circ \cO(0).
\end{align}
Similar cohomology equivalences apply to twisted special conformal transformations $\hg(\b,\bb)$.
Even though cohomology classes do not depend on either $\b$ or $\bb$, by \eqref{ghrels2.5}, these can still be important in correlators since they move world volumes and are needed to avoid intersections at space-like infinity.
We will use this to compute $\bS$-$\bS$ OPEs for the free hypermultiplet in section \ref{sec:4}.



\section{Convergence, $\TT_{\pm}$ Ward identities, and intersections}\label{sec:3}

Contributions to correlation functions from parallel time-, space-, or light-like lines in the world volumes of two or more descent operators may result in conditionally convergent expressions which require regularization.
$\TT_\pm$-boundaries refer to the boundary terms of \eqref{GD4} that must vanish in an appropriate sense for descent operators to be $\TT_\pm$-closed so that all $\TT_{\pm}$ Ward identities --- such as \eqref{E1} and those of section \ref{sec:2.5} --- are obeyed.
Formally, non-vanishing $\TT_\pm$-boundary terms can occur in correlator configurations with parallel time-, space-, or light-like world volume lines.
However, determining when these terms lead to violations of $\TT_{\pm}$ Ward identities is ambiguous unless integral expressions are absolutely convergent.
We find that there exists a rather tightly constrained regularization prescription which renders descent operator correlators absolutely convergent, and ensures all $\TT_\pm$ boundary terms vanish within $\TT_{\pm}$ Ward identities.
Therefore, such regulated descent operators represent well-defined twisted Schur cohomology classes, i.e., they yield unambiguous, finite results and automatically satisfy the $\TT_\pm$ Ward identities.
The specific properties of the regularization prescription for the hatted descent operators $\hbS_{\cdots}, \hbW^{\pm}_{\cdots}$ differ slightly from the prescription required for the unhatted descent operators $\bL^{\pm},\bS$.

This regulator is in addition to the usual Minkowski space-time regularization of propagator poles at light-like separations.
Since we require correlation functions to satisfy the twisted Schur cohomology Ward identities, e.g., \eqref{ghrels2} and \eqref{ghrels4}, we regulate light-cone poles using the usual $i\e$ prescription (reviewed below) so they produce time-ordered operator products in Minkowski space-time.

Local operator products or correlators must be finite only if all operator insertions avoid space-time intersections with each other, including at the space-time boundaries.
For local (Schur) operators $\cO_1(z_1)$, $\cO_2(z_2)$, this simply means that their VOA plane coordinates do not coincide, $z_1 \neq z_2$, but for descent operators this may not be sufficient to avoid world volume intersections.
Even after removing intersections between descent operators at finite points in space-time, they may still intersect at space-like, time-like, or light-like infinity.
These intersections are due to certain (sets of) parallel lines, and they generically occur between all descent operators world volumes except for $L^+$ relative to $L^-$.
For certain descent correlators, some of these intersections can be removed while preserving both $\TT_+$ and $\TT_-$ cohomology, or only one of them, depending on the cohomology properties of relevant conformal transformation.
In such cases, non-intersecting parallel lines might still remain that can lead to conditional convergence issues but, our proposed regularization will remedy these correlators.
Furthermore, for those intersections which cannot be removed with $\TT_{\pm}$-exact conformal transformations, the regularization prescription will still make such descent correlators absolutely convergent by removing any contribution intersecting points might have to the local correlator integrals.

For reasons we will describe, it remains to be proven that descent operator cohomology classes are independent of the choices involved in the regularization prescription. 

\subsection{Possible divergences in descent correlators}
\label{desc_div}

The basic defining properties of any $\TT_{\pm}$-closed operator must be that its insertion into a correlation function of twisted Schur cohomology classes in generic positions is finite, unambiguous, and obeys all $\TT_{\pm}$ Ward identities. 
After starting with an analysis of the asymptotic limits of a generic local operator 2-point function, we will identify a potential source of divergence within descent-descent 2-point functions that will serve as our primary motivation for introducing a regularization prescription in the next section, which will render these divergences finite and unambiguous.
We end by briefly describing a potential source for $\TT_{\pm}$ Ward identity violations that are ultimately cured by this regularization scheme.

Time-ordered correlators of descent operators are defined to be the world volume integrals of the time-ordered correlators of the local operators appearing in their integrands, e.g., 
\begin{align}\label{desc_corr}
    \Bigl\langle \int d\a_1 \cO_1(x_1(\a_1)) \, \int d\a_2 \cO_2(x_2(\a_2)) \Bigr\rangle &
    \doteq \int d\a_1 d\a_2 \ \bigl\langle \cO_1(x_1(\a_1)) \, \cO_2(x_2(\a_2)) \bigr\rangle.
\end{align}
The light-cone singularities of time-ordered correlators are regulated by the usual causal $i\e$ prescription where the time-difference coordinate $x_{12}^{\m=1} \doteq t_{12}$ is replaced by the complex value $(1+i\e) t_{12}$ and the limit $\e\to 0^+$ is taken.%
\footnote{Other $i\e$ prescriptions, such as $t_{12} \to t_{12} + i\e \,\sgn(t_{12})$ also work, though ones for which the imaginary shift goes to zero as $|t_{12}|\to\infty$ do not. 
The ``Wick rotation'' $t\to (1+i\e)t$ regularization seems more consonant with scale invariance.\label{iepftnt}} 

A given local conformal primary operator $\cO$, or any of its descendants%
\footnote{Replacing $\cO$ and $\bar\cO$ by descendants only improves the large-separation convergence of the 2-point function.} in the integrand can only give a non-vanishing contribution if there is an OPE channel involving its conformal conjugate field, $\bar \cO$, contributing a 2-point function 
\begin{align}\label{OLS3}
    \< \cO(x_1) \bar \cO(x_2) \> \sim ((x_{12})_{\b\bd})^{j+\tj} |x_{12}|^{-2(\D + \frac12(j +\tj))}
\end{align}
where $|x_{12}|^2 = (x_1-x_2)^\m (x_1-x_2)_\m$, $\D$ is the conformal dimension of $\cO$, and $(j,\tj)$ are its Lorentz spins.%
\footnote{We normalize our Lorentz spins to be integers, following the conventions of \cite{Cordova_2016}.}
$((x_{12})_{\b\bd})^{j+ \tj}$ is an appropriately symmetrized sum of $j + \tj$ powers of the matrix elements $(x_{12})_{\b\bd}$.
It follows from 4d unitarity that all operators (except the identity operator) obey $\D-\tfrac12(j+\tj)\ge1$.
The invariant distance-squared (with $i\e$ regularization) is
\begin{align}\label{propep}
        |x_{12}|^2 = -x_{12}^+x_{12}^- + |z_{12}|^2 
        - \tfrac{1}{2} i \e (x_{12}^+ + x_{12}^-)^2 
        + \tfrac{1}{4} \e^2 (x_{12}^+ + x_{12}^-)^2 .
\end{align}
This grows as $|x_{12}|^2 \sim (x_{12}^+)^2$ as $|x_{12}^+|\to\infty$, irrespective of how $x_{12}^-$ and $z_{12}$ behave in this limit.
The weakest possible fall-off of \eqref{OLS3} as $|x_{12}^+|\to\infty$ occurs when we take all spinor components in the numerator to be $(x_{12})_{1 \dot2} \sim x_{12}^+$, giving%
\footnote{For the more computationally convenient $i\e$ prescription, $t_{12} \to t_{12} + i \e \, \sgn t_{12}$, the fall-off of the 2-point function is weaker, $\lim_{|\a| \to \infty}  \< \cO(x_1) \bar \cO(x_2) \> \sim \a^{-\D+\frac12(j+\tj)}$, but still goes to zero by unitarity.}
\begin{align}\label{OLS3b}
    \lim_{|x_{12}^+| \to \infty}  \< \cO(x_1) \bar \cO(x_2) \> 
    \lesssim \lim_{|x_{12}^+| \to \infty} (x_{12}^+)^{-2\D} = 0.
\end{align}
An analogous inequality can be obtained from assessing the weakest convergence of \eqref{OLS3} in the asymptotic limit of any direction of space-time.

This analysis ensures the integrand of a 2-point function between a local Schur operator and a descent operator will have no divergences in the asymptotic world volume limits, and we find that the world volume integrals are absolutely convergent for finite $\e$. However, for a 2-point function $\<\cO(w)\bL^{\pm}[\cO']\>$ between an $\bL^{\pm}$ descent operator with a local Schur, $\cO(w)$, the result diverges in the $\e \to 0^+$ limit. Selection rules severely restrict when such 2-point functions can occur \cite{Argyres_2023xx}. 
    
Divergences can also appear in correlators with multiple descent operators.
To see this, let $f(x_{12}) \equiv \< \cO(x_1) \bar\cO(x_2)\>$ be the local operator integrand of a descent-descent 2-point function where the world volumes contain infinite parallel straight lines, $x_1=x_1(\a)$ and $x_2=x_2(\b)$ for $\a,\b\in\R$, such that $x_{12}=x_{12}(\a-\b)$. Then, $f(x_{12})$ is a function only of $\a-\b$.%
\footnote{This kinematic situation occurs if $\del_{\a} x_1(\a) = \del_{\b}x_2(\b)$ and both $x_1$ and $x_2$ can be written as $x_i^{\m} = t x_0^{\m} + c_i^{\m}$ for $t \in \R$ where $x_0^{\m}$ is a constant four-vector that is independent of any world volume parameters and $c_i^{\m}$ is a (potentially constant) four-vector containing remaining world volume parameters.\label{line_div}}
After performing the change of variables $\b \to \til \b + \a$, the world volume integrals factorize into the product of an integral of $f(x_{12})$ over $\til \b$ and other world volume parameters (which has no poles by the above estimates) times $\int d\a \cdot 1$, which diverges.

This type of world volume divergence is kinematically allowed in any correlators including a pair of descent operators with world volumes which are VOA plane translates of $L^+L^+$, $L^-L^-$, $L^\pm S$, $L^\pm \wh W$, $SS$, $S\wh W$, and $\wh W \wh W$.%
\footnote{This type of divergence doesn't occur in 2-point functions involving $\hbS$ world volumes because the light-like lines in such world volumes do not satisfy the properties described in footnote \ref{line_div}.}
Moreover, such a divergence can be argued to always occur in the special case that the integrands of the two descent operators in question are (derivatives of) conformally conjugate fields.%
\footnote{In a generic correlator containing two descent operators with parallel line world volumes and conformally conjugate integrands, we can apply the OPE between their integrands within the integration region of the correlator where points along these parallel lines are asymptotically separated from all other local operator insertion points in space-time. 
We can then apply the same reasoning which showed the 2-point functions of such descent operators are divergent.}
The pairs of descent operators with parallel lines in their world volumes and integrands that allow for conformally conjugate operators are tightly constrained by the pattern of super-descendants acting on a Schur primary that define a given descent operator \cite{Argyres_2023xx}.
For example, $\bL^+[\cO_1](z) \bL^+[\cO_2]$ and $\bL^+[\cO_1](z) \bS[\cO_2]$, though their world volumes contain infinite parallel lines, do not contain conjugate integrands no matter what Schur operators $\cO_j$ they are built from by descent.
Still, examples that can contain conjugate integrands do exist;  two simple cases are $\bS[\cO](z) \, \bS[\cO](0)$ and $\bL^+[\cO](z) \, \wh\bW^-_{\ldots}[\cO](0)$.

\paragraph{$\TT_\pm$ boundary terms.}

One might worry that, in addition to non-convergence of correlators of descent operators, they may violate the $\TT_\pm$ Ward identities by virtue of non-vanishing $\TT_\pm$ boundary terms.
In fact, such boundary terms only contribute to descent correlators which possess parallel line kinematics.%
\footnote{Non-vanishing boundary terms can appear in parallel line correlators regardless of whether or not the kinematics lead to the divergences previously described. This means they can occur for any descent operator, including the $\hbS_{\cdots}$ descent operators.}

To see this, consider the $\TT_+$ boundary limits \eqref{GD4} for $\bL^+$ and $\bS$ descent operators, 
which place operators at light-like infinity since they come from $P_+$-descent.%
\footnote{The $\TT_\pm$ boundary terms for the $\wh\bS_{\ldots}$ and $\wh\bW^\pm_{\ldots}$ descent operators are qualitatively different, and are discussed in section \ref{hat_conv}.} 
In particular, the $\TT_+$ boundaries contribute terms of the form $\lim_{|x^{+}|\to\infty} e^{x^+ P_+}\circ \cfE$ to correlator integrands where $\cfE$ is either a generic local operator $\cO(x_0)$ placed at an $x^+$-independent point $x_0$ or an $\bLm$ descent operator.
(``General local operator'' means not necessarily a twisted Schur operator.)
Analogous boundary limits are formed from the $\TT_-$ boundary terms of $\bLm$ and $\bS$.
When inserted into an arbitrary correlation function of local operators inserted at finite points in space-time, each boundary limit can only give a non-vanishing contribution if there is an OPE channel involving the local conformal conjugate operator to $\cfE$, $\bar \cO$, which contributes terms proportional to (derivatives of) the 2-point function \eqref{OLS3}.
Since the weakest possible fall-off of a 2-point function in any direction is given by \eqref{OLS3b}, this implies the $\TT_\pm$ boundary limits of $\bL^\pm$ or $\bS$ all vanish in correlators with local operators at finite points.
However, it does \emph{not} necessarily imply their vanishing in correlators with other descent operators.
This is because descent correlators integrate over local operator correlators that can contain multiple insertions on the conformal boundary of space-time for which the 2-point function doesn't vanish in the boundary limits.
For instance, consider the boundary limit, $\lim_{|x_1^+| \to \infty} e^{x_1^+P_+}\circ\cfE$, as an insertion in some multi-descent $\TT_+$ Ward identity, where $\bar\cO(x_2)$ is the integrand of another descent operator and the world volume integral over $x_2$ includes limits in which $x_2$ goes to space-time infinity.
From \eqref{propep}, it follows that when $x_2^+ = x_1^++c$ with $c$, $x_2^-$, and $z_2$ fixed, then $\lim_{|x_1^+|\to\infty} |x_{12}|^{-2} \neq 0$, so the 2-point function doesn't vanish everywhere for this choice of integration.
This corresponds to inserting $\bar\cO(x_2)$ at a point approaching light-like infinity along a direction parallel to an $L^+$ line and at the same rate as the boundary limit, $\lim_{|x_1^+| \to \infty} e^{x_1^+P_+}\circ\cfE$.
When a multi-descent correlator possesses the requisite kinematics, this potential non-vanishing of $\TT_\pm$ boundary terms within the world volume integrals of $\TT_{\pm}$ Ward identities could then lead to the Ward identities being violated.

Implicit in this argument were numerous choices on how to evaluate the boundary limits, $\lim_{|x_1^+| \to \infty} e^{x_1^+P_+}\circ\cfE$, relative to the evaluation of other world volume integrals. 
This can make the evaluation of $\TT_{\pm}$ boundary terms ambiguous since, in general, these different choices do not commute with one another to yield a unique answer unless the descent correlator was absolutely convergent to begin with. 
Furthermore, since these potential $\TT_{\pm}$ boundary violations occur in descent correlators with parallel line world volumes, it means such correlators will typically possess the divergences associated with such configurations as well.
Indeed, one can check that this formal non-vanishing of $\TT_{\pm}$ boundary terms can in correlators which suffer from such divergences.
These facts suggest that a regularization procedure which makes descent correlators absolutely convergent by regulating parallel line divergences, and additionally preserves $\TT_\pm$ Ward identities, will automatically be one which removes all contributions of the $\TT_\pm$ boundary limits. 
In this way, the evaluation of $\TT_{\pm}$ boundary terms in $\TT_{\pm}$ Ward identities becomes unambiguous, and any potential violations of such $\TT_{\pm}$ Ward identities are avoided.
We now describe a class of such regulators.

\subsection{Regularization of descent operators}\label{weight_func}

Regulate the descent operation \eqref{GD3} by adding a weight function, $F(\d,\a)$, to the integrand so,
\begin{align}\label{Fdescent}
\bD^X[F;\cfE] \doteq \lim_{\d\to 0^+}  \int_{-\infty}^\infty \!\!\! d\a \, F(\d,\a)\, 
\x \circ e^{\a X} \circ \cfE .
\end{align}
We demand $F(\d,\a)$ is differentiable for $\d>0$ and $\a\in\R$, and on this domain has the additional properties
\begin{align}\label{Fprops}
\text{(i)} & & |F(\d,\a)| & \le 1, \nn\\
\text{(ii)} & & \lim_{\d\to 0^+} F(\d,\a) & \equiv 1, \\
\text{(iii)} & & F(\d,\a) & < c(\d) \, |\a|^{-N} \qquad \text{for some $c(\d)>0$ and $N>1$,} \nn\\
\text{(iv)} & & \lim_{\d\to 0^+} \frac{\del F(\d,\a)}{\del\a} &\equiv 0. \nn
\end{align}
Property (i) means that the weight function can only improve the convergence of the descent integral; property (ii) ensures that the $\d\to 0^+$ limit coincides with the unregulated descent operation when the integral converges; (iii) means that $F$, and all of its $\a$-derivatives, suppress integrands by more than $|\a|^{-1}$ for large $|\a|$;  and (iv) ensures that the $\d\to 0^+$ limit produces an operator in $\TT_\pm$ cohomology.
In practice, the required value of $N$ is fixed by ensuring absolute convergence of the correlators for $\bD^{X}[F;\cfE]$.

Some simple weight functions which satisfy these requirements are
\begin{align}\label{FMdef}
F_M &\doteq (1+ i \d \a)^{-M}, \qquad 1 < M \in\Z, &
&\text{and} &
F_\infty &\doteq e^{-\d\a^2}.
\end{align}
For a given $F_M$, property (iii) is satisfied for all $N\leq M$ while, for $F_{\infty}$, it is satisfied for all $N<\infty$.
Note that while $F_\infty$ converges faster than any power of $|\a|$, it may not be as convenient as the $F_M$ in computations since it has an essential singularity at infinity in the complex $\a$ plane.

Also, note that the conditions (i)-(iv) proposed above for the regulator may not be the weakest possible ones.
For instance, condition (iii) could be replaced by the two slightly weaker conditions, $\lim_{|\a|\to\infty} |\a|^N F < \infty$ and $\lim_{|\a|\to\infty} \del_\a F/F < \infty$.
Ultimately, any property (iii) condition must be strong enough so that this regularization unambiguously renders all descent operators $\TT_{\pm}$-closed.
The following argument demonstrates that this is true for the conditions in \eqref{Fprops}, and it applies equally well for the slightly weaker version of property (iii).

Consider the descent by a $\TT_+$-exact generator, $\bD^{N_+}[\cfE]$, of some $\TT_\pm$ cohomology class  $\cfE$, regulated with weight function $F(\d,\a)$, 
\begin{align}
     \bD^{N_+}[F; \cfE] \deq \int d\a \, F(\d,\a)\, \n_+ \circ e^{\a N_+} \circ \cfE. 
\end{align}
We assume that $F$ obeys (iii) with sufficiently large $N$ so this integral is (absolutely) convergent.
For $P_\pm$ descent, which is all that's used for the $\cO$-$\bL^\pm$-$\bS$ subalgebra, the $\int d\a \cdot 1$ divergences appearing in correlators of descent operators with parallel line world volumes become $\int d\a F(\d, \a)$ which converge by property (iii) for any $N>1$. 
The same holds true for the $\wh\bS_{\cdots}$ and $\wh\bW^{\pm}_{\cdots}$ operators, but larger values of $N$ will be required; see the discussion in section \ref{hat_conv}.

Even with the regulator, we still have $\TT_- \circ \bD^{N_+}[F;\cfE] = 0$, since this follows from the conformal algebra without an integration by parts with respect to the descent parameter $\a$, so the presence of $F$ in the integrand is irrelevant.
But, the action of $\TT_+$ is now modified relative to \eqref{GD4},
\begin{align}\label{TpFDp}
   \TT_+ \circ \bD^{N_+} [F; \cfE] &= 
   \int d\a \, F(\d,\a) \, \del_\a  e^{\a N_+} \circ \cfE \nn \\
   &=  -  \int d\a \, \frac{dF(\d,\a)}{d\a} \, e^{\a N_+} \circ \cfE 
   + \Big[ F(\d,\a) e^{\a N_+} \circ \cfE \Big ]_{\a=-\infty}^{\a=+\infty} .
\end{align}
We now view \eqref{TpFDp} as an insertion in the local operator integrand of some $\TT_+$ Ward identity (which is by assumption absolutely convergent). By property (iii), the $\a$-derivatives of $F$ have increased convergence relative to $F$, so the first term will produce absolutely convergent integrals. Thus, we can take the $\d \to 0^+$ limit inside the integral which, by property (iv), means the first term vanishes in the limit.
The boundary limits $|\a| \to \infty$ involved in the second term both vanish for all $\d>0$ by property (iii).
This follows from our earlier analysis of $\TT_{\pm}$ boundary terms in section \ref{desc_div}.
By applying the local OPE to such an integrand, it is only non-zero if there is an OPE channel reducing to a 2-point function between $\cfE$ and its (local operator) conformal conjugate.
Then, this 2-point function can be at worst finite as $|\a|\to\infty$ for all world volume points of the correlator.
So, with the weight function regularization factor $F(\d,\a)$, the boundary terms in \eqref{TpFDp} vanish by (iii). 
Note that in arguing for this conclusion, it is understood that the regularization prescription requires one to evaluate any $\TT_{\pm}$ boundary limits \emph{before} taking the $\d \to 0^+$ limit.

Lastly, it is important to point out that this result implies all of the $\TT_{\pm}$ equivalence relations between descent operators that were derived in sections \ref{sec:2.4}-\ref{descent_web}. In those derivations, it was crucial that various boundary terms of descent operators vanished. Once weight functions are included in their definition and the $\d\to 0^+$ limits are appropriately taken, the previous argument shows that all of these boundary terms do indeed vanish.\\

We have thus shown that there exists a large family of regulators that render descent operator correlators convergent and preserve the $\TT_\pm$ Ward identities by ensuring that they are always $\TT_{\pm}$-closed.
However, we have not shown that the resulting descent operator $\TT_{\pm}$ cohomology classes are independent of the choice of regulator $F$.
Despite this, there are many correlators which are absolutely convergent without any regularization, and when regularized with any $F$ obeying (i)-(iv), they therefore give the same answers.
Furthermore, although we cannot convincingly prove that all results requiring this regularization are independent of the choice of weight function, at the end of section \ref{hat_conv}, we discuss preliminary evidence showing this regularization can indeed produce universal, i.e., regulator independent, results when it is required.

For the free hypermultiplet examples we compute in the next section, all the $\cO$-$\bL^\pm$-$\bS$ two-point functions are absolutely convergent (so do not need any regularization) with the exception of the $\< \bS \bS \>$ 2-point functions.
This is because the $\< \bS(z) \bS(0) \>$ 2-point functions have a further subtlety related to the fact that the two VOA plane translated world volumes intersect at space-like (and time-like) infinity.
So we turn now to this question of intersections at infinity.

\subsection{OPEs and intersections in the $\cO$-$\bL^\pm$-$\bS$ subalgebra}\label{L+-S_OPE+Int}

We compute operator products by inserting operators separated in space-time, and taking the limit as they approach one another.
In the case of $\cO(z)$, $\bL^\pm[\cO](z)$, and $\bS[\cO](z)$, if their $z$ coordinates are different their world volumes do not intersect at any finite points in space-time.
However, since their world volumes extend to infinity, we must be careful when considering their intersections at points at infinity.
For this, it is useful to work in conformally compactified space-time, which is the familiar Penrose diamond with boundaries the future and past light-like infinities $\cfI^\pm \simeq I \times S^2$ together with future and past time-like infinity points $i^\pm$ and a space-like infinity point $i^0$.
Then $\bL^\pm(z)$ and $\bL^\mp(0)$ do not intersect at infinity, while $\bL^\pm(z)$ and $\bL^\pm(0)$ or $\bS(0)$ intersect at 2 points on $\cfI^\pm$, and $\bS(z)$ and $\bS(0)$ intersect at 4 points on $\cfI^\pm$ and at $i^0$ and $i^\pm$.
The intersections at light-like infinity of $\bL^\pm$ with $\bL^\pm$ or $\bS$ turn out to be innocuous when computing their OPEs with the $i\e$ prescription.
This is because none of these products can close on the identity operator, which means a 2-point function can't exist between them so no world volume integrals of the corresponding propagators occur.
By contrast, the $\bS$-$\bS$ OPE closes on the identity, so the $\bS$ with $\bS$ intersections at $\cfI^\pm$ and at $i^0$ and $i^\pm$ give rise to divergences which must be further regulated.

One might expect that since parallel space-like or time-like lines can be conformally transformed into non-parallel configurations, that they do not give rise to the parallel line divergences discussed in section \ref{desc_div}.
However, they are, in fact, transformed into intersecting configurations, so possible divergences persist, and the intersections must be removed to put the correlator into a generic non-intersecting configuration.
By contrast, parallel light-like lines remain parallel light-like lines intersecting at $\cfI^\pm$ under general conformal transformations.
Thus, intersections at space-like or time-like infinity are qualitatively different from those at light-like infinity.

We start with the intersections at the space-like point at infinity, $i^0$, and the past and future time-like infinity points, $i^\pm$.
The space-like infinity point is also the point at infinity on the VOA plane.
Just as in previous discussions of VOA modules associated to surface operators \cite{Cordova:2017, Bianchi_2019}, a surface operator insertion twist-translated to $z$ on the VOA plane should also be thought of as providing an insertion at the point at infinity.

The twist special conformal transformations of the VOA plane, $\hg(\b)$ in \eqref{ghrels1}, move the points at space-like and time-like infinity to finite points.
While the $\bS(z)\, \bS(0)$ operator product has intersections at $i^0$ and $i^\pm$, the $\bS(z) \, \hg(\b){\circ}\bS(0)$ product where just one of the insertions is acted on by a twist special conformal transformation, does not have any such intersections for $\b\neq0$.
The first cohomology equivalence in \eqref{ghrels4} indicates that the operator product should not depend on $\b$, but the topological distinction between intersections at $i^{0,\pm}$ when $\b=0$ and no intersections at these points when $\b\neq0$ makes it possible that there are two different ($\b$-independent) answers in these two cases.
This is generically the case for any $\bS$-$\bS$ 2-point function. 
For example, for the scalar $\D_\cO=1$ Schur primaries $\cO=q_I$ of the free hypermultiplet SCFT discussed in section \ref{sec:4}, we find that $\langle \bS[q_I](z) \, \hg(\b){\circ}\bS[q_J](0) \rangle = - \e_{IJ} z^{-1}$ for $\b\neq0$, but vanishes for $\b=0$ by a selection rule.\footnote{Typically, when no selection rules prevents such a 2-point function, a $\b$-independent $\bS$-$\bS$ 2-point function is formally divergent due to parallel line divergences.}
When computing all $\bS$-$\bS$ products, in addition to the weight function regularization, we regulate these intersections at space- and time-like infinity using the $\hg(\b)$ transformations.

In computing these $\b$-regulated products, one must take into account that the $\hg(\b)$ transformations act discontinuously on Minkowski space-time since they are special conformal transformations.
The natural way to compute these products is on the Lorentzian cylinder, which is covered by an infinite number of copies of conformally compactified Minkowski space-time, and on which conformal transformations act continuously \cite{Penrose:1964}.
Also, time-ordered Lorentzian cylinder correlators of unitary CFTs are the analytic continuation of correlators in the CFT on the euclidean 4-sphere \cite{Luscher:1974}.

But these correlators suffer from a second intersection problem.
Turning on a relative twist conformal transformation, $\b\neq0$, does not remove the intersections at the four points on $\cfI^\pm$ at light-like infinity.
This is because $\hg(\b)$ transformations fix the $L^\pm$ lines.
Indeed, there are no relative conformal transformations of the two $S$ world volumes that both preserve twisted Schur cohomology and also remove these intersection points.
In fact, these non-removable parallel light-like lines in the two $S$ surfaces are the source of the divergence of the $\langle \bS \bS \rangle$ 2-point function which is proportional to the volume of $S$.
Due to this, one must separately regulate the $\bS$ descent operators using weight functions, as discussed in section \ref{weight_func}.

However, the calculation of the 2-point function when the surface operators have both a relative twist-translation, a relative twist-special conformal transformation, and a weight function regulator is difficult.
We bypass this difficulty for the purposes of the calculations presented in section \ref{sec:4} by ``Wick rotating" $\bS$-$\bS$ 2-point functions to euclidean space.
There are no light-like infinities in Euclidean space:  there is just a single point, $i^0$, at (spatial) infinity compactifying Euclidean space to $S^4$.
A relative (euclidean) $\hg(\b)$ special conformal transformation separates the intersection of two 2-surfaces at $i^0$, and completely regulates their 2-point function.
Mathematically, this Euclidean regularization is easy, but comes at the cost of not having a physically satisfactory interpretation in terms of extended operator world volumes in Minkowski space-time or analytically continued Wightman functions. 
Because of this, a primary future goal of ours is to compute $\bS$-$\bS$ 2-point functions using the weight function regularization presented in section \ref{weight_func}, with the intention of determining if the answers are independent of the particular weight function chosen.


\paragraph{Configuration space discontinuities, intersections, and $\TT_{\pm}$ Ward identities.} 
The fact that the 2-point functions $\<\bS(z)\, \hg(\b)\circ \bS(0)\>$ and $\<\bS(z) \bS(0)\>$ disagree in a discontinuous way indicates that multi-descent correlators can have discontinuities based on their world volume configurations.
For these 2-point functions, the discontinuity appears after removing intersections at $i^0$ and $i^{\pm}$ using a $\hg(\b)$ twist special conformal transformation.
Such discontinuities can be viewed as existing within the cohomology configuration space of the descent correlator.
This space parameterizes the cohomology-allowed, non-intersecting, world volume configurations of the operator insertions, or rather, the allowed $\TT_{\pm}$-closed transformations that can act on such insertions.%
\footnote{Cohomology configuration spaces of a descent operator can be viewed as coset spaces constructed from the superconformal group where, physically, they parameterize the different points/world volumes of a local/descent operator cohomology class. 
When working within twisted-Schur cohomology, these spaces are drastically simplified for each descent operator. $n$-point descent correlators can be viewed as being valued in (products of) certain quotients of descent operator cohomology configuration spaces and, locally about any regular point, they look like $n$ copies of $\C_z$. 
This is equivalent to the statement that the twist-translated correlators of the vertex algebra only depend on their $z$-insertion points within the VOA plane.}
In this language, we can rephrase the disagreement between $\bS$-$\bS$ 2-point functions as the statement that the $\bS$-$\bS$ 2-point function has an intersection discontinuity in the component of its configuration space that parameterizes the twist special conformal transformations, $\hg(\b)$.
This means the discontinuity separates the configuration space into two regions: one where $\b=0$ and the correlator is zero or divergent, and one where $\b \neq 0$ and the correlator is finite.
For any correlator with intersections at $i^0$ and $i^{\pm}$, we expect it to have this $\hg(\b)$ intersection discontinuity.

The hatted descent operators can also have intersection discontinuities.
For example, they will occur in $\<\hbS_{\cdots}\hbS_{\cdots}\>$ 2-point functions but, in this case, the associated intersections typically occur at finite points in space-time. 
These discontinuities occur in the components of the $\<\hbS_{\cdots}\hbS_{\cdots}\>$ configuration space that parameterize the twist translations, $\hg(z)$, and twist rotations, $\hg(i\phi)$, of \eqref{ghrels1}.
However, it is also true that every correlator of a hatted descent operator contains $\hg(z)$ and $\hg(i\phi)$ discontinuities that are \emph{not} intersection discontinuities.
This is the case for, e.g., the 2-point function between a free hyper Schur, $q_I$, and its corresponding $\hbS_{\cdots}$ descent operator.
These discontinuities are inherent to the $M_{\pm3}$ descent procedure, and their interplay is discussed in section \ref{hat_OPE+Ints}.
Since every wall operator includes an $M_{\pm3}$ descent procedure, we expect all of their correlators to possess the non-intersection, $\hg(z)$ and $\hg(i\phi)$ discontinuities.
For example, these discontinuities will appear in the 2-point function between a wall operator and an $\bL^{\pm}$ line operator.
In addition to these non-intersection, $M_{\pm 3}$ discontinuities, we also expect the correlators of wall operators to have $\hg(\b)$ intersection discontinuities when they possess intersections at $i^0$ and $i^{\pm}$. 
For instance, this would be true for the 2-point, $\<\hbW^{\pm}_{\cdots}\hbW^{\pm}_{\cdots}\>$, between wall operators. 

All of these configuration space discontinuities indicate that the $\TT_{\pm}$ Ward identities leading to the twist transformation equivalence relations, \eqref{ghrels2.5}, are violated for certain pairs of descent operators that are related by a twist transformation but, are also separated by a discontinuity of some kind.
For physically reasonable correlators, the intersection discontinuities are allowed to exist. 
However, the non-intersection discontinuities possessed by vertex algebra correlators is a new feature that vertex operator algebra correlators --- constructed from only local Schur operators --- do not possess.
Instead of being an inconsistency, this appears to be an allowed feature of descent correlators that is due to their extended world volumes, and the fact that the twist transformations are composed of the $\sl_2$ symmetry generators, which are $\TT_{\pm}$-closed and \emph{not} $\TT_{\pm}$-exact.%
\footnote{Formally, a twisted Schur cohomology correlator is allowed to depend on the transformations of any symmetry generator that is $\TT_{\pm}$-closed and not $\TT_{\pm}$-exact.}
For this reason, we do not expect there to exist discontinuities on the components of a descent correlator configuration space that parameterize the real, $\TT_{\pm}$-exact transformations generated by $\fm_{\pm}$.
For all correlators and OPEs we've computed, including those that require the weight function regularization, they remain invariant under all $\fm_{\pm}$ transformations.

This discussion also makes clear why, even though the light-like intersections on $\cfI^{\pm}$ in certain descent correlators can be removed using $\fm_{\pm}$ transformations, they will not change the value of such a correlator.
Hence, there will not exist intersection discontinuities within the $\fm_{\pm}$ components of such correlator configuration spaces. 
As discussed in section \ref{weight_func}, the remedy for these correlators is to employ the weight function regularization.%
\footnote{Invariance under all $\fm_{\pm}$ transformations serves as a constraint on the weight functions. 
Based on the discussion in section \ref{weight_func}, this is automatically satisfied since the weight function properties guarantee any such regulated descent operator is $\TT_{\pm}$-closed.}

\subsection{$\TT_{\pm}$ boundaries and regularization of $\hbS_{\cdots}$ and $\hbW^{\pm}_{\cdots}$ descent operators}
\label{hat_conv}

The construction of any $\hbS_{\ldots}$ or $\hbW^\pm_{\ldots}$ descent operator involves a single $M_{\pm3}$ descent procedure.
The $M_{\pm3}$ generators act on the Lorentz representation of a Schur operator at the origin and, unlike the orbits of the $P_\pm$ translation generators, the infinite-parameter boundaries of their space-time orbits include finite points in space-time.
These facts make the $\hbS_{\cdots}$ and $\hbW^\pm_{\cdots}$ operators less convergent than the $\bL^\pm$ and $\bS$ operators.
We will describe the convergence properties of $M_{\pm3}$ descent using $\hbS_R$ which will motivate the precise regularization prescription required to make any hatted descent operator a well-defined $\TT_{\pm}$ cohomology class. Similar arguments will hold for $\hbS_L$ and the $\hbW^{\pm}_{\ldots}$ operators.

Using its definition \eqref{hbSLR}, write the fully regulated $\hbS_R[F_R; F_+;\cO]$ as the double descent integral 
\begin{align}
    \hbS_R[F_R;F_+;\cO] = \lim_{\d_R \to 0^+} \lim_{\d_+ \to 0^+} \int_{\R^2} d\xi \, dv \,  F_R F_+ \eta_{-3} \circ 
    e^{-\xi M_{-3}} \circ \pi_+ \circ e^{v P_+} \circ \cO(0),
\end{align}
where $F_+(\d_+,v)$, $F_R(\d_R,\xi)$ are the weight function regulators for the $P_+$ and $M_{-3}$ descent procedures, respectively, $\pi_+ \deq -i \Qt_{1\dot1}$ is the descent supercharge satisfying ${\TT_+ \circ \pi_+ = P_+}$, and $\eta_{-3} \deq - \frac{i}{2} (Q^1_{~ 1} + \St^{2 \dot1})$ is the descent supercharge satisfying $\TT_- \circ \eta_{-3} = M_{-3}$.%
\footnote{
Here $Q^a_{\ \a}$, $\Qt_{a\ad}$, $S_a^{\ \a}$, and $\St^{a\ad}$ are the supertranslation and superconformal generators of the $\cN{=}2$ superconformal algebra in the notation of \cite{Beem_2015, Lemos_2020}.  
In particular, $\a$, $\ad$, $a$ are Lorentz left and right spinor indices and $\SU(2)_R$ doublet indices, respectively.}
$M_{-3}$ is a non-compact generator formed from a light-like combination of a rotation and a boost which generates a subgroup of elements $\{g(\x) \doteq e^{-\x M_{-3}},\, \x\in(-\infty,\infty)\}$.
Acting on the light-like $L^+$ line, $g(\x)$ has the effect of rotating $L^+$ to other light-like lines in the light cone $\hS$ shown in figure \ref{fig1}, reaching the $L^-$ line only in the $\x \to \pm\infty$ limit.
Thus, the $\x\to\pm\infty$ boundaries of the $\bD^{M_{-3}}$ descent operation include finite points in space-time, namely all the points on $L^-$.
Parameterize the $\hS$ light cone by the ``cone" coordinates $x^\m(t,\th)$ with $(x^1,x^2,x^3,x^4) = t (1,\cos\th,\sin\th,0)$ where $t\in\R$ and $\th\in[-\pi,\pi]$.
Note that $\th=0$ is $L^+$ and $\th=\pm\pi$ is $L^-$.

In order for $\hbS_R[F_R;F_+;\cO]$ to be well-defined as a cohomology class, the weight functions $F_+$ and $F_R$ must each satisfy certain convergence properties which are constrained by $N_+$ and $N_R$, respectively, as described in property (iii) of \eqref{Fprops}. 
To fix lower bounds on $N_+$ and $N_R$, we examine the convergence of the $\TT_\pm$ boundary terms of $\hbS_R[F_R;F_+;\cO]$.%
\footnote{We cannot fix the lower bound on $N_+$ from parallel line divergences as we did for the unhatted descent operators because the $\hbS_{\cdots}$ descent operators do not experience those kinematic divergences.}
These have different divergence behaviors depending on the Lorentz spins of the local Schur operator. 
The $\TT_+$ boundary term gives the most severe divergences and will therefore provide the strongest convergence requirements on the weight functions $F_R$ and $F_+$.
Acting with $\TT_+$ on $\hbS_R[F_R;F_+;\cO]$ yields a single boundary term at $t = \pm \infty$,
\begin{align}\label{T+_ShR_bdry}
\TT_+ \circ & \hbS_R[F_R;F_+;\cO] = \frac12 \int_{-\pi}^{\pi} d\th \, F_R \sec^2(\th/2) \ 
\Bigl[ F_+\,  e^{x^{\m}(t,\th)P_{\m}} \ \circ  \\
& \qquad \circ  \left ( i (Q^1_{\ 1} + \St^{2 \dot1}) + t \left (i \sin \th \, Q^2_{\ 1} + (1+ \cos \th) Q^2_{\ 2}\right) \right ) \circ e^{i \x \cMt^{\dot1}_{\ \dot2}} \circ \cO(0) \Bigr]_{t=-\infty}^{t=+\infty}, \nn 
\end{align}
where $F_R \equiv F_R(\d_R, \th)$ and $F_+\equiv F_+(\d_+,t,\th)$ for $v = t(1+ \cos \th)$, $\x = \tan (\th /2)$. 
Regardless of the order in which the $t$-limits and $\th$-integral are evaluated, the boundary limits of \eqref{T+_ShR_bdry} are required to vanish.
There are two sources of divergence in \eqref{T+_ShR_bdry} that occur in the limits $\th \to \pm \pi$ of the $\th$-integral. First, the $\th$-integral has a second order pole at $\th = \pm \pi$ due to the $\sec^2(\th/2)$ factor.
Second, if the Schur $\cO$ is a component of a field with Lorentz spins $(j,\tj)$ then, 
\begin{align}\label{Shthbdry2}
    e^{i\x \cMt^{\dot1}_{\ \dot2}} \circ \cO(0) \sim 
    \cO_{\dot1\cdots\dot1}(0) + \cdots + \x^\tj \cO_{\dot2\cdots\dot2}(0),
\end{align}
where we show only the $\tj$ dotted spinor indices of $\cO$.
Since $\xi \to \pm \infty$ as $\th \to \pm \pi$, these terms increase the divergence of $\TT_+ \circ \hbS_R[F_R;F_+;\cO]$. In \cite{Argyres_2023xx} we will explicitly confirm that this operator has divergent 2-point functions for the free hypermultiplet.
The conclusion is that, generically, the unregulated descent operator $\hbS_R[\cO]$ will not be unambiguously $\TT_+$-closed for any Schur operator $\cO$.

A regularization of these divergences is achieved using the same prescription introduced in section \ref{weight_func} once the weight functions $F_+$ and $F_R$ satisfy property (iii) in \eqref{Fprops} for specific values of $N_+$ and $N_R$.
In particular, the value of $N_R$ will depend on the Lorentz representation of the Schur operator that the hatted descent operator is constructed from. 
For $\hbS_R[F_R;F_+;\cO]$, the convergence of \eqref{T+_ShR_bdry} indicates, 
\begin{align}\label{SR_(iii)}
\hbS_R[F_R;F_+;\cO]:  \quad N_{R} > 1 + \tj \qand N_{+} > 2 \qq{for property (iii)},
\end{align}
where the Schur $\cO$ has Lorentz spins $(j,\tj)$.
The value of $N_R$ is obtained by requiring the $\th$-integral to be integrable near $\th = \pm \pi$, and the value of $N_+$ is obtained by requiring the integrand of \eqref{T+_ShR_bdry} to converge to zero in the $t \to \pm \infty$ limits.%
\footnote{This value of $N_+$ ensures the integrand of a non-zero 2-point function involving $\TT_+\circ \hbS_R[F_R;F_+;\cO]$ and another extended operator will always vanish in the $t \to \pm \infty$ limits even when the local operator 2-point function doesn't converge to zero because parallel line kinematics prevent $|x_{12}|^{-2}$ from converging to zero.}
When this is satisfied, all divergences that could occur in both $\TT_{\pm}\circ \hbS_R[F_R;F_+;\cO]$ are regulated. 
Furthermore, the arguments in section \ref{weight_func} following \eqref{TpFDp} prove that all $\TT_{\pm}$ boundary terms of $\hbS_R[F_R;F_+;\cO]$ will vanish in all $\TT_{\pm}$~Ward identities. 
A mirror analysis of the $\TT_-\circ \hbS_R[F_L;F_-;\cO]$ boundary term gives
\begin{align}\label{SL_(iii)}
\hbS_R[F_L;F_-;\cO]:  \quad N_{L} > 1 + j \qand N_{-} > 2 \qq{for property (iii)},
\end{align}
where here, $N_L$ and $N_-$ are the property (iii) parameters bounding the weight functions $F_L$ and $F_-$ corresponding to the $M_{+3}$ and $P_-$ descent procedures of $\hbS_L$, respectively.

The $\TT_{\pm}$ boundary terms for the $\hbW^{\pm}_{\cdots}$ operators can be examined in a similar fashion.
We will present the resulting property (iii) convergence bounds on their weight function regulators in \cite{Argyres_2023xx}.\\

Unlike the $P_{\pm}$ weight functions of $\bL^{\pm}$ and $\bS$ descent operators, the $M_{\pm3}$ weight functions of a hatted operator are needed in every correlator they're involved in, even when all other insertions are local Schur operators. 
This is because there are intrinsic divergences associated with the $M_{\pm3}$ descent procedure, and these do not depend on relative world volume kinematics. 
Instead, they stem from the geometry of the $\hS$ and $\hW$ world volumes, and the infinite boost limits of their Lorentz transformations on a local Schur operator, $\cO(0)$.

These $M_{\pm3}$ divergences may not always appear in a twist-translated correlator of a hatted descent operator, or in any twist transformation of such a correlator. 
For instance, the twist-translated 2-point function, $\< q_I(z) \hbS_R[F_R;q_J](0)\>$, built from the Schur operator of the free hyper presented in section \ref{sec:4}, is absolutely convergent and obeys the twist-translate Ward identity without the weight function $F_R$.%
\footnote{We don't show the weight function $F_+$ associated to the $P_+$ descent procedure of $\hbS_R$ because it has no effect on this correlator, or any $\TT_{\pm}$ cohomology transformations of it.}
However, these divergences will generically appear whenever a hatted correlator is transformed by a $\TT_{\pm}$-exact transformation from the $\fm_{\pm}$ subalgebras of \eqref{Tcohom2}. 
This means correlators of hatted descent operators will not obey the $\fm_{\pm}$ Ward identities unless they are equipped with the weight function regularization associated to the $M_{\pm3}$ descent procedure.

For the 2-point function $\< q_I(z) \hbS_R[F_R;q_J](0)\>$, where $F_R(\d_R,\xi)$ is a \emph{any} weight function satisfying \eqref{Fprops} and \eqref{SR_(iii)}, we can prove the following:
\begin{outline}
    \1 All integrals involved in $\fm_{\pm}$ transformations of it are absolutely convergent;
    \1 It is invariant under the action of any $\fm_{\pm}$ transformation.
\end{outline}
Together, these results provide promising support for the claim that descent operator cohomology classes are independent of the choice of weight function regulator that is used to define them. 
We will present these results in \cite{Argyres_2023xx}.

\subsection{OPEs and intersections of $\hbS_{\cdots}$ and $\hbW^{\pm}_{\cdots}$ descent operators}\label{hat_OPE+Ints}

The $\hS$ and $\wh W$ world volumes break the rotational invariance of the VOA plane by picking out the $z=\zb$ direction.
Acting on these world volumes with the twist transformation, $\hg(\ell)$ of \eqref{ghrels1} for imaginary $\ell = i \phi$, produces VOA-plane rotated world volumes which we denote by $\hS_\f$ and $\hW_\f$.
Here, $\f$ represents the angle from the real $z$-axis in the VOA plane, so the $\hS_\f$ world volume is thus extended in the $z=e^{2i\f} \zb$ direction.
The corresponding descent operators, $\hbS_{\cdots}(\f) \deq \hg(i\phi)\circ \hbS_{\cdots}$ and  $\hbW^{\pm}_{\cdots}(\phi) \deq \hg(i\phi)\circ \hbW^{\pm}_{\cdots}$, are generated by descent using $M_{L}(\phi) \doteq \cos\f M_{+3} + \sin\f M_{+4}$ for the ``$L$" descent operators or $M_{R}(\phi) \doteq \cos\f M_{-3} + \sin\f M_{-4}$ for the ``$R$" descent operators.
The $\hg(\ell)$ cohomology equivalence of \eqref{ghrels4} shows that these $\hg(i\phi)$-transformed descent operators satisfy the equivalence relations,%
\footnote{Note that when these equivalences hold, they also hold in the presence of weight function regulators. For this reason, we do not always write out the weight functions explicitly in this section.}
\begin{align}\label{hat_ghl_equiv}
   \hbS_{\cdots}[\cO](\f) &=_\TT \hbS_{\cdots}[\cO], &
    \hbW^{\pm}_{\cdots}[\cO](\phi) &=_\TT \hbW^\pm_{\cdots}[\cO] .
\end{align}
Thus, formally, there seems to be nothing special about the $z=\zb$ direction in the VOA plane in the extended vertex algebra.

However, as mentioned at the end of section \ref{L+-S_OPE+Int}, we find that the correlators of $\hbS_{\cdots}(\phi)$ and $\hbW^{\pm}_{\cdots}(\phi)$ operators will generically experience non-intersection discontinuities across certain lines in the VOA plane, even in their 2-point functions with local Schur operators.%
\footnote{The existence of such discontinuities in vertex algebra correlators is in stark contrast to the correlators of the vertex operator algebra that are constructed from twist-translations of local Schur operators. 
At non-intersecting points, the latter correlators are guaranteed to be meromorphic and, therefore, have no branch cuts or discontinuities in them.}
This means the equivalence relations in \eqref{hat_ghl_equiv} will only hold within certain regions of the VOA plane whose boundaries are these lines of discontinuity, which depend on $\f$ and the twist-translation parameter(s), $z$.
Therefore, such discontinuities break the equivalence between, say, $\hbS_{\cdots}[\cO](\f)$ and $\hbS_{\cdots}[\cO]$, for a generic value of $\f$.

As an example, the 2-point functions $\< q_I(z)\hbS_R[F_R;q_J](\phi,0)\>$ contain a factor of, $\sgn(\Re (e^{i\phi }z))$, which characterize these lines discontinuity, and the regions within which the equivalence \eqref{hat_ghl_equiv} holds.
In particular, the equivalence ${\hbS_{\cdots}(\phi)[\cO] =_{\TT} \hbS_{\cdots}[\cO]}$ will only hold in the region of $(\phi,z)$-configuration space where $\sgn(\Re (e^{i\phi }z)) = \sgn(\Re(z))$.
Within this region, these discontinuities are ``quasi-topological" in the sense that they can be rotated using a $\hg(i\phi)$ transformation without changing the value of the 2-point function.

Since these discontinuities only appear in descent operators built from the Lorentz boosts, $M_L(\phi)$ and $M_R(\phi)$, we will refer to them as ``boost discontinuities".
We expect boost discontinuities to appear in all correlators of a hatted descent operator and, when they do, they will restrict the region of $(\phi,z)$-configuration space where the equivalences in \eqref{hat_ghl_equiv} hold.%
\footnote{This is very similar to what happens for the $\<\bS\bS\>$ 2-point functions discussed in section \ref{L+-S_OPE+Int}. There, the $\hg(\b)$ equivalence relation of \eqref{ghrels2.5} only holds in $\bS$-$\bS$ 2-point functions that have already been $\hg(\b)$-transformed, i.e., $\<\bS(z)\bS(\b)\> = \<\bS(z) \hg(\b') \circ \bS(\b)\>$ only if $\b \neq 0$. Here, we are saying that the $\hg(i \phi')$ equivalence relation of \eqref{ghrels2.5} only holds within the 2-point function $\<q_I(z)\hbS_R[q_J](\phi,0)\>$ if $\sgn(\Re(e^{i\phi}z )) = \sgn(\Re(e^{i(\phi+\phi')}z))$. }

Generically, the world volumes of an $\hbS_{\cdots}(\f,0)$ operator and a twist-translated $\hbS_{\cdots}(\f',z)$ operator do not intersect at any finite points in space-time.
The exception is the specific arrangement where ${\rm arg}(z) = (\f+\f'+\pi)/2$.
Thus, correlators $\langle \hbS_{\cdots}(\f_1,z_1) \hbS_{\cdots}(\f_2,z_2) \cdots \rangle$ will have regions in their $(\f_i, z_i)$-configuration space where the cohomological equivalence in \eqref{hat_ghl_equiv} is satisfied.
The boundaries of these regions correspond to intersection discontinuities in the $(\phi_i,z_i)$-configuration space, and points on the boundary are the excluded configurations where pairs of $\hS_{\f}$ world volumes intersect in space-time.
The geometry and topology of these intersection discontinuities within $(\phi,z)$-configuration spaces is complicated, and we defer further study of them to \cite{Argyres_2023xx}.

Once we insert a $\hbW^{\pm}_{\cdots}(\phi)$ operator in a correlator, then any other $\hbS_{\cdots}(\f')$ or $\hbW^{\pm}_{\cdots}(\phi')$ insertions must have parallel world volumes, i.e., all must have $\f'=\f+n\pi$.
This is simply because, if their world volumes are not parallel in this sense, they will intersect the $\hbW^{\pm}_{\cdots}(\phi)$ insertion at finite points in space-time.

Lastly, the $\hbW^{\pm}_{\cdots}(\phi)$ world volume intersects the VOA plane along the $z=e^{2i\f}\zb$ line.
Thus, we expect correlators of $\hbW_{\cdots}^{\pm}(\phi)$ operators to have intersection discontinuities along this line of intersection with the VOA plane.%
\footnote{There are also possible intersection discontinuities coming from the world volume intersection of a wall operator with the points $i^0$ and $i^{\pm}$ of space- and time-like infinity.} 
Within a correlator containing a $\hbW^{\pm}_{\cdots}(\phi)$ operator, its boost and VOA-plane intersection discontinuities will always lie orthogonal to one another within the VOA plane.
For this reason, the VOA-plane intersection discontinuities will also be quasi-topological within certain components of the $(\phi,z)$-configuration space.
This means a $\hg(i\phi)$ rotation of the intersection line in the VOA plane will not change a $\hbW^{\pm}_{\cdots}(\phi)$ correlator in cohomology until the rotation moves the intersection line across the VOA-plane insertion point of another vertex operator.
The effect is that correlators of a $\hbW^{\pm}_{\cdots}(\phi)$ operator are only well-defined if all other vertex algebra insertions avoid this line.


The existence of boost and VOA-plane intersection discontinuities means the hatted descent operators are not, strictly speaking, vertex operators. 
Instead, their existence suggests these operators behave more like boundary or line operators in the VOA plane.
This characterization applies in a literal sense to the $\hbW^\pm_{\cdots}(\phi)$ operators since their world volumes physically intersect the VOA plane along a line.


\section{Vertex algebra of the free hypermultiplet theory}\label{sec:4}

The set of vertex operators (at the origin of the VOA plane, say) containing Schur operators $\cO(0)$ and their corresponding descent operators, $\bL^{\pm}[\cO]$ and $\bS[\cO]$, is characterized by having support in the $z=\zb=0$ plane in space-time, and so, in particular, does not contain the $\hbS_{\ldots}$ or $\hbW^\pm_{\ldots}$ descent operators.
We now calculate some operator products in this $\cO$-$\bL^\pm$-$\bS$ sub-vertex algebra within the free hypermultiplet SCFT.  
In particular, we will calculate the OPEs among the free dimension-1 scalar Schur, $q$, the dimension-3 vector Schur operator, $T$, and their $\bL^\pm$ and $\bS$ descent operators.
$T$ is the Virasoro operator in the VOA which is the chiral theory of a free symplectic boson.

Since the fermionic super generators $\pi_\pm$ satisfying $\TT_\pm \circ \pi_\pm = P_\pm$ are
\begin{align}\label{OLS1}
    \pi_+ &= -i \Qt_{1\dot1}, &
    \pi_- &= -i Q^2_{\ 1},
\end{align}
it follows from the definition \eqref{Ndesc1} of the descent operation that
\begin{align}\label{OLS2}
\bLp[\cO] &= -i N \int \!\! dx^+ \,
\Qt_{1\dot1} \circ e^{x^+ P_+}\circ \cO(0), 
\hspace{10mm}
\bLm[\cO] = -i N \int \!\! dx^- \,
Q^2_{\ 1} \circ e^{x^- P_-}\circ \cO(0), \nn\\
\bS[\cO] &= - N^2 \int \!\! dx^+ dx^- \, Q^2_{\ 1} \circ \Qt_{1\dot1} \circ  e^{x^+ P_+ + x^- P_-}\circ \cO(0),
\end{align}
Their normalizations are a matter of choice; we have put in the factors of $N$ for later convenience.
These descent operators are especially easy to work with since $\pi_\pm$ and $P_\pm$ all commute.
$\bL^\pm[\cO]$ are reminiscent%
\footnote{The $\bL^\pm[\cO]$ descent lines are not quite primary light-ray transforms.
The natural generalization of the light-ray transform to a conformal primary with general Lorentz spins $(j,\tj)$ is $\bL[\cO][\z,\til\z] = \int ds \, \cO_{\a_1 \cdots \a_j\, \ad_1 \cdots \ad_\tj}(s \z^\a \til\z^\ad) \ \z^{\a_1} \cdots \z^{\a_j} \,  \til\z^{\ad_1} \cdots \til\z^{\ad_\tj}$, where $\cO$ is integrated over the lightline $x^\m \s_\m^{\a\ad} = s \z^\a \til\z^\ad$ where $\z^\a$ and $\til\z^\ad$ are polarization spinors such that $x^\m$ is real.  
The $\bL^\pm[\cO]$ light-like line operators are not of this form: their spinor polarizations do not line up with the $L^\pm$ world lines.
This means they have transverse spin relative to the null-plane, and should therefore be interpreted as a kind of ``primary descendant" of the corresponding primary light-ray transform constructed from the Schur operator they are built from. We thank David Simons-Duffin for explaining this point to us.}
of the light-ray transform operators introduced in \cite{Kravchuk_2018:1805.00098}, and some of the discussion of \cite{Kravchuk_2018:1805.00098, Kologlu:2019, Chang:2020} on the computation of their correlators and OPEs applies. 

The free hypermultiplet superconformal primaries are the free complex scalars, $\r^a_I$, satisfying the reality conditions $(\r^a_I)^\dag = \e^{IJ} \e_{ab} \r^b_J$, where $a,b \in \{1,2\}$ are fundamental $\su(2)_R$ indices and $I,J \in \{ 1, 2\}$ are fundamental $\su(2)_F$ flavor indices.%
\footnote{All $\e$ symbols satisfy $\e_{12}=-\e^{12}=1$.}
There are also free fermions given by $\ps_{\a I} \doteq \frac12 \e_{ba} Q^a_{\ \a} \r^b_I$ and $\pst_{\ad I} \doteq \frac12 \Qt_{a \ad} \r^a_I$.
Their 2-point functions are
\begin{align} \label{} 
    \langle \r^a_I(x_1) \r^b_J(x_2) \rangle 
    &= \frac{\e^{ab} \e_{IJ}}{|x_{12}|^2} ,&
    \langle \ps_{\a I}(x_1) \pst_{\ad J}(x_2)\rangle 
    &= \frac{\e_{IJ} (x_{12})_{\a \ad}}{|x_{12}|^4} .
\end{align}
where
\begin{align}
x_{\a\ad} \doteq x^\m \s_{\m\a\ad} 
&= \bpm \zb & -ix^+  \\ ix^- & -z \epm . 
\end{align}

We consider the twist-translated Schur operators 
\begin{align}\label{qTdef}
    q_I(z) &\doteq u_a \r^a_I(z,\zb), & 
    T(z) &\doteq u_a u_b \tfrac12 \e^{IJ} \nord{\r^a_I\del_z \r^b_J}(z,\zb) ,
\end{align}
where $u_a = (1, - \zb)$.
In the VOA, $q_I$ is a free symplectic boson of chiral weight $h_q = 1/2$, and $T$ is the Virasoro operator of weight $h_T = 2$.


\paragraph{$\bL^{\pm}$ and $\bS$ descent operators.} Using the definitions of the descent operators given in section \ref{sec:2}, the twist translated descent operators for $q_I$ are
\begin{align} \label{qDescentOps} 
\bLp[q_I](z) &= -i N \int_{L^+}dx^+ \, \pst_{\dot 1I}(x^+,0,z,\zb),
\nn\\
\bLm[q_I](z) &= -i N \int_{L^-}dx^- \, \ps_{1I}(0,x^-,z,\zb), \\
\bS[q_I](z) &= - N^2\int_{S}dx^+ dx^- \, \del_z \r^2_I(x^+,x^-,z,\zb)
\nn
\end{align}
where we choose the normalization factor
\begin{align}
    N = (-2\pi i)^{-1/2},
\end{align}
to simplify the form of the OPEs. 
The twist translations act trivially on these descent operators because the local operators being integrated are all $\su(2)_R$ singlets or of lowest $\su(2)_R$ weight. 
This is not the case for the twist translated descent operators of $T$ which are
\begin{align} \label{TDescentOps} 
\bLp[T](z) &= \frac{-i}{2} \e^{IJ} N \int_{L^+} dx^+\, 
u_a \left ( \nord{\pst_{\dot 1 I} \del_z \r^a_J} 
+ \nord{\r^a_I\del_z \pst_{\dot 1J}} \right ) (x^+,0,z,\zb), \nn\\
\bLm[T](z) &= \frac{-i}{2} \e^{IJ} N \int_{L^-}dx^- \, 
u_a \left ( \nord{\ps_{1I}\del_z \r^a_J} 
+ \nord{\r^a_I\del_z\ps_{1J}} \right )(0,x^-,z,\zb), \\
\bS[T](z) &= -\frac{\e^{IJ}}{2} N^2 \int_{S} dx^+ dx^- \, 
\Big\{u_a  (\nord{\del_z\r^2_I\del_z\r^a_J} 
+ \nord{\r^a_I\del_z\del_z\r^2_J})(x^+,x^-,z,\zb) \nn \\
& \hspace{45mm} \mbox{}
+ (\nord{\ps_{1I}\del_z\pst_{\dot 1J}} 
- \nord{\pst_{\dot 1I}\del_z\ps_{1J}})(x^+,x^-,z,\zb)
\Big\}.\nn
\end{align}

\paragraph{$\cO$-$\bL^{\pm}$-$\bS$ OPEs.} For the OPEs involving a descent operator, one must often compute integrals of propagator-like expressions to simplify the result. 
Whenever possible we perform such integrals in Minkowski space-time using the standard $i\e$ prescription but, as is the case with 2-point functions of $\bS$, this is not always sufficient. 
Curiously, in cases where there are space-time intersections and the computation of integrals are involved,  we have checked that the $i \e$ prescription is sufficient to compute them, as long as these integrals come from terms not closing on the identity.

An important property of these OPEs is that they are meromorphic functions of $z$ as required by cohomology Ward identities.
This is a check that we are computing descent operator products correctly.

A priori, additional $\TT_{\pm}$-closed extended operators might appear on the right sides of the OPEs which are not equivalent in cohomology to the descent operators of $q_I$ and $T$. 
This would indicate we were ``missing" distinct extended operators in the $\cO$-$\bL^\pm$-$\bS$ subset which are not necessarily captured by descent. 
In practice, seemingly distinct $\TT_{\pm}$-closed surface/line operators indeed appear in the OPEs involving a descent operator. 
However, whenever this occurs, we have always found that subtle integral cohomology identities end up showing these seemingly new extended operators are in fact $\TT_\pm$-equivalent to the descent operators of $q_I$ and $T$ or their normal-ordered products. 
We detail these cohomology identities in \cite{Argyres_2023xx}.

In the end, we find that the OPEs among these Schur operators and their descent operators can all be put into canonical forms which we specify in a condensed notation with the following tables.\\

\noindent\textbf{$q_I,\, \bL^{\pm}[q_I]$ and $\bS[q_I]$ OPEs:}
\begin{gather}
X_I(z) Y_J(0) \sim \, \e_{IJ} \frac{\tred{a}}{z}
\qfor \tred{a} \in \R \nn
\end{gather}
\vspace{-2em}
\renewcommand{\arraystretch}{1.5}%
\begin{align}\label{qqOPE}
\begin{array}{c|cccc}
X_I \backslash Y_J &  q_J & \bS[q_J ] & \bLp[q_J ] & \bLm[q_J] \\  \hline 
q_I & -1 & -1 & 0 & 0 \\
\bS[q_I] & -1 &-1 & 0 & 0 \\
\bLp[q_I] & 0 & 0 & 0 & -1 \\
\bLm[q_I] & 0 & 0 & 1 & 0 
\end{array}
\end{align}
The coefficient of the $\bS[q_I]\bS[q_J]$ OPE comes from performing the integrals involved in Euclidean space; see the discussion in section \ref{sec:3}.\\

\noindent\textbf{$T, \, \bL^{\pm}[T], \bS[T]$ and $q_I,\, \bL^{\pm}[q_I],\bS[q_I]$ OPEs:} 
\begin{gather}
X_{T}(z) Y_I(0) \sim \, \left(\frac{1}{2z^2} + \frac{1}{z} \del_z \right ) \tred{V_I}(0) 
\qfor \tred{V_I} \in \{ q_I, \, \bL^{\pm}[q_I], \, \bS[q_I] \} \nn
\end{gather}
\begin{align}\label{TqOPE}
\begin{array}{c|cccc}
  X_T \backslash Y_I & q_I & \bS[q_I] & \bLp[q_I] & \bLm[q_I] \\  \hline 
T & q_I & q_I & 0 & 0 \\
\bS[T] & \bS[q_I] &  \bS[q_I] & \bLp[q_I] &  \bLm[q_I] \\
\bLp[T] & \bLp[q_I] & \bLp[q_I] & 0 & q_I \\
\bLm[T] & \bLm[q_I] & \bLm[q_I] & q_I & 0 
\end{array}
\end{align}

\vspace{1mm}
\noindent\textbf{$T, \bL^{\pm}[T], \bS[T]$ OPEs:}
\begin{align}\label{XTYTOPE}
X_T(z) Y_T(0) &\sim  - \frac{\tred{a}}{2z^4} 
+ \frac{2 (\trU\trV)}{z^3} + \frac{\del_z(\trU\trV)}{z^2} \ \, +\frac{\del_z^2(\trU\trV)}{z}  \nn\\
& \hspace{35mm}+ \frac{2(\trU{\wedge}\trV')}{z^2} + \frac{\del_z(\trU{\wedge}\trV')}{z} \nn\\
& \hspace{58mm} + \frac{(\trU'\trV')}{z} \\
& \equiv \, (\tred{a},\trU,\trV)
\qfor \tred{a} \in \R \qand \trU, \trV \in \{q, \, \bL^{\pm}[q], \, \bS[q] \}, \nn
\end{align}
where we have defined the normal ordered products
\begin{align}\label{UVno}
    (UV) & \doteq -\frac{\e^{IJ}}4 
    \nord{U_I V_J}(0) \ , \nn\\
    (U{\wedge}V') & \doteq +\frac{\e^{IJ}}4 
    \nord{(U_I \,\del_z V_J - \del_z U_I \, V_J )}(0) \ , \\
    (U'V') & \doteq \e^{IJ}
    \nord{\del_z U_I \,\del_z V_J}(0) \ . \nn
\end{align}
Then
\begin{align}\label{TTOPE}
\begin{array}{c|cccc}
  X_T \backslash Y_T & T & \bS[T] & \bLp[T] & \bLm[T] \\  \hline 
T & (1, q, q) & (1, q, \bS[q]) & (0, q, \bLp[q]) & (0, q, \bLm[q]) \\
\bS[T] & (1, \bS[q], q) &  (1,\bS[q],\bS[q]) & (0, \bS[q], \bLp[q]) & (0, \bS[q], \bLm[q]) \\
\bLp[T] & (0,\bLp[q],q) & (0,\bLp[q],\bS[q]) & (0, \bLp[q], \bLp[q]) & (1, \bLp[q], \bLm[q]) \\
\bLm[T] & (0,\bLm[q],q) & (0,\bLm[q],\bS[q]) & (-1, \bLm[q], \bLp[q]) & (0, \bLm[q], \bLm[q])
\end{array}
\end{align}
Note that 
\begin{align}
    (UV) &= -(-)^{|U|\cdot|V|} (VU), &
    (U{\wedge}V') &= +(-)^{|U|\cdot|V|} (V{\wedge}U'), &
    (U'V') &= -(-)^{|U|\cdot|V|} (V'U'), \nn
\end{align}
where $|U|=0$ for a boson, and $|U|=1$ for a fermion.

\paragraph{Remarks and observations.} We now make a few preliminary observations on the structure of this operator algebra.
The most striking observation is that while the $\bL^\pm[q]$ OPEs in \eqref{qqOPE} describe a pair of conjugate free complex fermions, upon diagonalizing the bosonic 2-point functions \eqref{qqOPE} by defining the new basis (suppressing flavor indices)
\begin{align}\label{diag1}
    q_\pm \doteq \frac12(q \pm \bS[q]),
\end{align}
their OPEs, in the notation of \eqref{qqOPE}, become
\begin{align}\label{qqOPE2}
\begin{array}{c|cc}
X \backslash Y &  q_- & q_+ \\ \hline 
q_- & 0 & 0 \\
q_+ & 0 & -1 
\end{array}  \qquad .
\end{align}
This means that $q_+$ is a free symplectic boson. 
But $q_-$ has a completely regular OPE, analogous to a chiral ring element.
In fact, it is immediate from \eqref{qqOPE} and \eqref{TqOPE} that $q_-$ has completely regular operator products with \emph{all} the fields we have constructed.

This raises the question of whether $q_-$ vanishes in cohomology: does
\begin{align}\label{question1}
    \bS[q_I] =_\TT q_I  \qquad ?
\end{align}
Clearly, such an equivalence would correspond to a VA-type equivalence.
We do not see a direct way of proving such a relation just on the basis of their expressions as operator representatives of twisted Schur cohomology classes in the 4d SCFT.
An indirect way is to show that all their operator products coincide with a generating set of operators for the extended vertex algebra.
The VOA of the free hyper SCFT is the free symplectic boson that is (strongly) generated by $q_I$, but we do not know of a generating set of operators for the extended vertex algebra, even though the underlying 4d SCFT is free.
Indeed, determining a generating set requires understanding the cohomological equivalences among descent operators and their normal-ordered products.
Note that even if the equivalence in \eqref{question1} between a local operator and its surface descent operator were true, this would have to be a property particular to $q_I$.  
For instance, we must have $T \neq_\TT \bS[T]$ since their operator products with $\bL^\pm[q_I]$ shown in \eqref{TqOPE} are different.

Another observation is that, as expected, the $T$ vertex operator has the usual Virasoro OPE with central charge $-1$,
\begin{align}\label{VOA1}
    T(z) T(0) \sim \frac{-1}{2z^4} + \frac{2 T(0)}{z^2} + \frac{\del_z T(0)}{z} ,
\end{align}
by virtue of the fact that $T=(q{\wedge}q') = \frac12 \e^{IJ} q_I \del_z q_J$ by \eqref{qTdef}.
But although $T$ is a Virasoro operator, it is not the stress-energy operator of the vertex algebra.
This is because to be a stress-energy operator of a vertex algebra, the algebra must be decomposable into a basis of primary operators, $\Phi_i(z)$, and their descendants, $\del_z^n\Phi_i(z)$, such that
\begin{align}\label{VOA2}
    T(z) \Phi_i(0) \sim \frac{h_i\Phi_i(0)}{z^2} + \frac{\del_z\Phi_i(0)}{z} ,
\end{align}
for some $h_i\in\N/2$.
If a stress-energy operator exists obeying \eqref{VOA1} and \eqref{VOA2}, then the vertex algebra is called a \emph{conformal vertex algebra} \cite{kac1998vertex} or, somewhat confusingly, a \emph{vertex operator algebra}.
Define the modes of $T$ by the expansion $T(z) = \sum_n T_n z^{-n-2}$ in the usual way.
The conformal vertex algebra conditions imply that the action of these modes on vertex operators coincides with that of the $L_n$  $\sl_2$ generators, i.e., $T_n=L_n$ for $n\in\{-1,0,1\}$.
In particular, $T_{-1}=\del_z$ and the eigenvalues of $T_0$ are the chiral weights $h_i$.
The chiral weight of $T(z)$ is $h_T=2$.

In our case $T$ does not obey \eqref{VOA2}, so it is not a stress-energy operator for the extended vertex algebra.
This is clear from \eqref{TqOPE} since $T(z) \bS[q](0) \sim (\cdots) q(0)$.
A key question is then whether there exists a stress-energy operator for our extended vertex algebra.
We look for one by searching for suitable bosonic operators with chiral weight 2.
Recalling that $q$ and its descent operators $\bL^\pm[q]$ and $\bS[q]$ all have weight $h_q=1/2$, while $T$ and its descent operators all have weight $h_T=2$, the possible weight-2 bosonic operators appearing in the OPEs we have calculated are
\begin{align}\label{VOA3}
    (q{\wedge}q'), \quad
    (q{\wedge}\bS[q]') , \quad
    (\bS[q]{\wedge}\bS[q]'), \quad
    T, \quad \bS[T] ,\quad
    (\bLp[q]{\wedge}\bLm[q]') ,
\end{align}
in the notation of \eqref{UVno}.%
\footnote{Note that $\bLp[q]\wedge \bLp[q]'=0$ and similarly for $\bLm[q]$.}
We have already remarked that $T=(q{\wedge}q')$.
If \eqref{question1} were true, then the first four operators of \eqref{VOA3} would be equivalent. 
Equivalences in $\TT_\pm$ cohomology among the remaining weight-2 operators, though possible, are not apparent.
We find no linear combination of these operators which satisfies both \eqref{VOA1} and \eqref{VOA2} for the subset of operators we are examining, even if we assume \eqref{question1} is true.
This, unfortunately, does not settle the question of whether the extended vertex algebra is a conformal vertex algebra since we have only looked at a limited number of operators and their OPEs.%
\footnote{For example, we did not take into account the various dimension two flavor singlets that come from flavor contractions between the moment map operator, $\mu_{IJ} \sim \nord{q_Iq_J}$, of the flavor multiplet for the free hyper and its corresponding descent operators. Examples of these Sugawara-like operators are, $\nord{\bX[q_I]\bY[q_J]q^I q^J}$ or $\nord{\bX[\nord{q_Iq_J}]q^I q^J}$, where $\bX, \bY \in \{ \bL^{\pm}, \bS\}$. We will report on these questions and the general properties of the flavor multiplet for the free hyper in \cite{Argyres_2023xx}.}

We end this section by making some more general remarks about the computation of descent operator OPEs which are motivated by the following questions,%
\footnote{We thank the referees for prompting us to answer these.}
\begin{outline}
    \1 What $4d$ local operator OPE data contributes to the descent operator OPEs that give rise to the vertex algebra?
    \1 Is this local operator OPE data determined by the OPEs of Schur operators which compose the VOA?
\end{outline}
It is easiest to answer these questions by distinguishing between free and interacting theories, and we will first talk about the former.
Recall from \eqref{desc_corr} that a time-ordered descent correlator is defined to be the world volume integral of the time-ordered correlator of its integrand.
For descent operator OPEs, a similar definition will hold only if (in every correlation function) we can apply the OPE channel between their local operator integrands for all pairs of points on their relative world volumes.
In a free theory, where all local correlators are reduced to two-point functions via Wick contractions, we believe that this is true because the OPE radius of convergence between any two local operators does not depend on the location of other local operator insertions in the correlator.%
\footnote{It is possible new $\TT_{\pm}$-closed extended operators may appear in this process, but the free field argument implies their integrand OPEs will still be determined by $4d$ local operator data.
From the results of the free hypermultiplet we've presented, we expect that it will always be possible to rewrite these operators in terms of descent operators and their $\del_z$ descendants.}
Thus, the OPEs of free field descent operator integrands are entirely determined in terms of the OPEs of their integrands, which are ordered combinations of certain super-descendants of a (twist-translated) Schur operator that are simultaneously composed with correlated $\fm_{\pm}$ transformations (see \eqref{GD2}-\eqref{GD3}).
The OPE coefficients of these super-descendants represent the additional $4d$ local operator data that contributes to free field descent correlators and OPEs.
However, note that there is no sense in which the result of performing world volume integrals can be determined by local operator data --- these are independent computations and their results are predicted beforehand by the $\TT_{\pm}$ Ward identities that descent operators satisfy.

It's well known that the OPEs of primary operators in a superconformal multiplet are not determined in terms of the OPEs of the corresponding superconformal primary \cite{Fortin:2011nq}.
Hence, we do not expect the OPEs of the integrands of free field descent OPEs to be determined by the VOA OPEs of Schur operators, and we have not found this to be the case for the OPEs we've computed for the free hypermultiplet in this section.
Furthermore, in the simplest (non-zero) descent operator OPE, $\bLp[q_I](z)\bLm[q_J](0)$, one can immediately see that the OPE of their integrands cannot be obtained from the VOA OPE, $q_I(z) q_J(0)$, using cohomological arguments because the former involves transformations using relative $P_{\pm}$ translations which break simultaneous $\TT_{\pm}$-cohomology.

In interacting SCFTs, the radius of convergence between local operators becomes finite \cite{Pappadopulo:2012jk}.
This implies it is not possible to compute the OPEs of descent operators in a generic VA correlator as we did for free fields, because the OPE channel between their integrands does not (generically) converge for all pairs of points on their relative world volumes.
This means interacting descent operator OPEs cannot be partially determined by the full set of local operator data of the SCFT in any direct way, if at all.
Instead, following the OPE work on light-ray operators (e.g. \cite{Kologlu:2019, Chang:2020}), one can resort to a more abstract approach where the OPEs of descent operators are inferred from their insertion into 4-point functions containing (at least) two other Schur operators.%
\footnote{We do not know if these 4-point functions are sufficient to deduce the full OPE of descent operators --- it is possible we must consider more general 4-point functions containing mixtures of descent operators insertions as well.}
Because they can be applied in general SCFTs, these more abstract methods can serve as a check on the simpler free field approach previously described.%
\footnote{Checking the free hypermultiplet and vector multiplet results should inform us of the necessary 4-point functions one needs to consider to apply the abstract approach.} \\
Because one can't rely on local operator data to compute them, it is possible that interacting descent OPEs contain additional information that goes beyond the $4d$ OPE data of local operators, which would be an expression of the fact that they are truly non-local operators.
Furthermore, this also means it is possible that interacting descent OPEs will contain new $\TT_{\pm}$-closed extended operators that can't be rewritten in terms of the descent operators we've constructed. 
Because we used free field methods to obtain the OPEs among $T$ and its descent operators in \eqref{XTYTOPE}-\eqref{TTOPE}, this discussion makes it clear that these results cannot be extrapolated to the VOA stress tensor for a generic interacting SCFT --- additional data and extended operators may be needed to reproduce such OPEs.


\section{Future directions}\label{sec:5}

We have shown how a version of topological descent applied to twist-translated Schur operators gives rise to a large set of new protected line, surface, and wall operators, which enlarges the vertex operator algebra of any 4d $\cN{=}2$ SCFT.
Some details of this construction were deferred to \cite{Argyres_2023xx}.

A computational hurdle is to formulate an appropriate regularization of descent correlators since, as discussed in section \ref{sec:3}, they generically contain divergences.
For the $\<\bS\bS\>$ 2-point functions presented in section \ref{sec:4}, we sidestepped this problem by computing them in euclidean signature.
In section \ref{sec:3}, we proposed a regularization prescription for all descent operators that involves the inclusion of a weight function in their definition.
These weight functions render descent correlator integrals absolutely convergent and their cohomology classes are well-defined.

An outstanding question regarding this proposed regularization is whether or not the resulting cohomology classes depend on the specific choice of weight function. 
At the end of section \ref{hat_conv}, we gave preliminary evidence indicating descent operator cohomology classes are independent of this choice.
In \cite{Argyres_2023xx}, we will analyze this question more thoroughly by reporting on whether the regulated $\bS$-$\bS$ 2-point functions are unique and weight function independent.
Showing this would provide strong evidence that our proposed regularization will produce universal results for all descent operators and, hence, for the extended vertex algebra.

Additional tasks are understanding the analytic structure of correlators involving the $\hbS_{\cdots}$ and $\hbW^\pm_{\cdots}$ quasi-topological operators, and understanding the 4d-type equivalences among the wall operators, $\hbW^\pm_{\cdots}$.

A key question is whether the extended vertex algebra is a conformal vertex algebra (a.k.a., a vertex operator algebra).  
A number of open questions were also raised in section \ref{sec:4} concerning the structure of the $\cO$-$\bL^\pm$-$\bS$ subalgebra.
The computation of the OPEs recorded in section \ref{sec:4} relied crucially on various series of cohomological operator identities, whose derivation is given in \cite{Argyres_2023xx}.
To sort out these central structural questions it seems necessary to find a way of systematically organizing and generating these identities. 

Extended vertex algebra OPE computations in the free vector multiplet SCFT is a next obvious step.
By combining free hypermultiplet and vector multiplet results, weak coupling extended vertex algebras in conformal gauge theories can also be obtained.
It would be interesting to study extended vertex algebra modules built on light-like Wilson-'t Hooft-type line operators in twisted Schur cohomology.
These should also be computationally accessible in conformal gauge theories.

More broadly, we would like to identify the ways in which the extended vertex algebra differs from the vertex operator algebra composed of Schur operators.
For instance, in \cite{Beem_2015} it was shown using the non-renormalization results of \cite{Baggio:2012rr} that the VOA of Schur operators is independent of exactly marginal couplings.
It is not clear if the required arguments for this result continue to hold for the descent operator correlators of the VA.

\acknowledgments

It is a pleasure to thank 
C. Beem,
L. Bianchi,
A. Bourget,
M. Bullimore,
M. Dedushenko,
C. Elliott,
A. Ferrari,
I. Garcia-Etxebarria,
J. Grimminger,
O. Gwilliam,
P. Kravchuk,
M. Lemos,
M. Martone,
L. Rastelli,
and D. Simmons-Duffin
for helpful discussions and suggestions.
PCA and MW are supported in part by DOE grant DE-SC0011784.
ML is supported in part by the National Research Foundation of Korea (NRF) Grant 2021R1A2C2012350.


\bibliographystyle{JHEP}
\bibliography{EVA_lett-V3.bib}

\end{document}